\documentclass[twocolumn]{aastex61}
\usepackage{CJK}
\usepackage{amsmath}
\usepackage{verbatim}

\newcommand\aastex{AAS\TeX}

\definecolor{blue}{RGB}{50, 80, 255}

\received{2019 March 8}
\revised{2019 April 15}
\accepted{2019 April 19}

\submitjournal{ApJ}

\shorttitle{\aastex\ Shallow Ultraviolet Transits of WD~1145+017}
\shortauthors{S. Xu et al.}

\begin{document}
\begin{CJK}{UTF8}{gbsn}
\title{Shallow Ultraviolet Transits of WD~1145+017}

\correspondingauthor{Siyi Xu}
\email{sxu@gemini.edu}

\author[0000-0002-8808-4282]{Siyi Xu (许\CJKfamily{bsmi}偲\CJKfamily{gbsn}艺)}
\affil{Gemini Observatory, 670 N. A'ohoku Place, Hilo, HI 96720}

\author[0000-0002-0430-7793]{Na'ama Hallakoun}
\affiliation{School of Physics and Astronomy, Tel-Aviv University, Tel-Aviv 6997801, Israel}

\author{Bruce Gary}
\affiliation{Hereford Arizona Observatory, Hereford, AZ 85615, USA}

\author[0000-0002-4297-5506]{Paul A. Dalba}
\affiliation{Department of Earth \& Planetary Sciences, University of California Riverside, 900 University Ave, Riverside CA 92521 USA}

\author{John Debes}
\affiliation{Space Telescope Science Institute, Baltimore, MD 21218, USA}

\author{Patrick Dufour}
\affiliation{Institut de Recherche sur les Exoplan$\grave{e}$tes (iREx), Universit$\acute{e}$ de Montr$\acute{e}$al, Montr$\acute{e}$al, QC H3C 3J7, Canada}

\author{Maude Fortin-Archambault}
\affiliation{Institut de Recherche sur les Exoplan$\grave{e}$tes (iREx), Universit$\acute{e}$ de Montr$\acute{e}$al, Montr$\acute{e}$al, QC H3C 3J7, Canada}

\author[0000-0002-4909-5763]{Akihiko Fukui}
\affiliation{Department of Earth and Planetary Science, Graduate School of Science, The University of Tokyo, 7-3-1 Hongo, Bunkyo-ku, Tokyo 113-0033, Japan}
\affiliation{Instituto de Astrof\'isica de Canarias, V\'ia L\'actea s/n, E-38205 La Laguna, Tenerife, Spain}

\author{Michael A. Jura}
\altaffiliation{Deceased}
\affiliation{Department of Physics and Astronomy, University of California, Los Angeles, CA 90095-1562, USA}

\author{Beth Klein}
\affiliation{Department of Physics and Astronomy, University of California, Los Angeles, CA 90095-1562, USA}

\author[0000-0001-9194-1268]{Nobuhiko Kusakabe}
\affiliation{Astrobiology Center,  2-21-1 Osawa, Mitaka, Tokyo 181-8588, Japan}

\author[0000-0002-0638-8822]{Philip S. Muirhead}
\affiliation{Department of Astronomy \& Institute for Astrophysical Research, Boston University, 725 Commonwealth Avenue, Boston, MA 02215, USA}

\author[0000-0001-8511-2981]{Norio Narita (成田\CJKfamily{bsmi}憲保)}
\affiliation{Department of Astronomy, The University of Tokyo, 7-3-1 Hongo, Bunkyo-ku, Tokyo 113-0033, Japan}
\affiliation{Astrobiology Center, 2-21-1 Osawa, Mitaka, Tokyo 181-8588, Japan}
\affiliation{JST, PRESTO, 7-3-1 Hongo, Bunkyo-ku, Tokyo 113-0033, Japan}
\affiliation{National Astronomical Observatory of Japan, 2-21-1 Osawa, Mitaka, Tokyo 181-8588, Japan}
\affiliation{Instituto de Astrof\'isica de Canarias, V\'ia L\'actea s/n, E-38205 La Laguna, Tenerife, Spain}

\author{Amy Steele}
\affiliation{Department of Astronomy, 1113 Physical Sciences Complex, Bldg. 415, University of Maryland, College Park, MD 20742-2421, USA}

\author[0000-0002-3532-5580]{Kate Y. L. Su}
\affiliation{Steward Observatory, University of Arizona, 933 N. Cherry Avenue, Tucson, AZ 85721, USA}

\author{Andrew Vanderburg}
\affiliation{Department of Astronomy, The University of Texas at Austin, 2515 Speedway, Stop C1400, Austin, TX 78712}

\author[0000-0002-7522-8195]{Noriharu Watanabe}
\affiliation{Optical and Infrared Astronomy Division, National Astronomical Observatory, Mitaka, Tokyo 181-8588, Japan}
\affiliation{Department of Astronomical Science, Graduate University for Advanced Studies (SOKENDAI), Mitaka, Tokyo 181-8588, Japan }

\author[0000-0002-4142-1800]{Zhuchang Zhan (詹筑畅)}
\affiliation{Department of Earth, Atmospheric and Planetary Sciences, Massachusetts Institute of Technology, Cambridge, MA 02139, USA}

\author{Ben Zuckerman}
\affiliation{Department of Physics and Astronomy, University of California, Los Angeles, CA 90095-1562, USA}

\begin{abstract}

WD~1145+017 is a unique white dwarf system that has a heavily polluted atmosphere, an infrared excess from a 
dust disk, 
numerous broad absorption lines from circumstellar gas, and changing transit features, likely from fragments of an actively 
disintegrating asteroid. Here, we present results from a large photometric and spectroscopic campaign with {\it 
Hubble}, {\it Keck }, VLT, {\it Spitzer}, and many other smaller telescopes from 2015 to 2018. Somewhat 
surprisingly, but consistent with previous observations in the u' band, the UV transit depths are always 
shallower than those in the optical. We develop a model 
that can quantitatively explain the observed ``bluing" and the main findings are: I. the transiting objects, 
circumstellar gas, and white dwarf are all aligned along our line of sight; II.  the transiting object is 
blocking a larger fraction of the circumstellar gas than of the white dwarf itself. Because most circumstellar 
lines are concentrated in the UV, the UV flux appears to be less blocked compared to the optical during a transit, leading to a shallower UV 
transit. This scenario is further supported by the strong anti-correlation between optical transit depth and 
circumstellar line strength. We have yet to detect any wavelength-dependent transits caused by the transiting material around WD~1145+017.

\end{abstract}

\keywords{circumstellar matter -- minor planets, asteroids: general -- stars: individual: WD 1145+017 -- white dwarfs}

\section{Introduction} \label{sec:intro}

There is evidence that planetary systems can be common and active around white dwarfs 
\citep[e.g.][]{JuraYoung2014, Veras2016}. To-date, WD~1145+017 is the only white dwarf that shows transit 
features of planetary material, likely from an actively disintegrating asteroid \citep{Vanderburg2015}. The 
original {\it K2} light curves reveal at least six stable periods, all between 4.5-5.0 hours, near the 
white dwarf tidal radius. Follow-up photometric observations show that the system is actively evolving and the 
light curve changes on a daily basis \citep{Gaensicke2016,Rappaport2016,Rappaport2017,Gary2017}. Likely, the 
transits are caused by dusty fragments\footnote{In this paper, we use the term ``dusty fragments" to refer to the material that is directly causing the transits. Likely, all these dusty fragments come from one or several asteroid parent bodies in orbit around WD~1145+017. But the asteroid itself is too small to be directly detectable via transits.} coming off the disintegrating asteroid \citep{Veras2017} and each piece is actively producing dust for a few weeks to many months.

The basic parameters of WD~1145+017 are listed in Table~\ref{tab:par}. Its photosphere
is also heavily ``polluted" with elements heavier than 
helium; such pollution has been observed in 25-50\% of all white dwarfs 
\citep{Zuckerman2003,Zuckerman2010,Koester2014a}. At least 11 heavy elements have been detected in its 
atmosphere and the overall composition resembles that of the bulk Earth \citep{Xu2016}. In addition, 
WD~1145+017 displays strong infrared excess from a dust disk, which has been observed around 40 other white 
dwarfs \citep{Farihi2016}. The standard model is that these disks are a result of tidal disruption of 
extrasolar asteroids and atmospheric pollution comes from accretion of the circumstellar material 
\citep{Jura2003}.

Another unique feature of WD~1145+017 is its ubiquitous broad circumstellar absorption lines, which have not been 
detected around any other white dwarfs \citep{Xu2016}\footnote{Variable circumstellar emission features have 
been detected around some dusty white dwarfs \citep[e.g.][]{Gaensicke2006,Manser2016,Dennihy2018}.}. They are 
broad (line widths $\sim$~300~km~s$^{-1}$), asymmetric, and arise mostly from transitions with lower energy 
levels $<$3 eV above ground. 
The circumstellar lines display short-term variability 
-- a reduction of absorption flux coinciding with the transit feature \citep{Redfield2017, Izquierdo2018,Karjalainen2019}, 
as well as long-term variability -- they have evolved from being strongly red-shifted to blue-shifted in a 
couple of years \citep{Cauley2018}. The long-term variability can be explained by precession of an eccentric 
ring, either under general relativity or by an external perturber.

\begin{deluxetable}{lllllll}
\tablecaption{Basic Parameters of WD~1145+017 \label{tab:par}}
\tablewidth{0pt}
\tablehead{
\colhead{Parameter} & \colhead{}  & \colhead{Ref}
}
\startdata
Coord (J2000)	& 11:48:33.6 +01:28:59.4	\\
Spectral Type	& DBZA	\\
V	&	17.0 mag\\
T$_\mathrm{WD}$	&15020 $\pm$ 520 K	& (1)	\\
log g	&	8.07 $\pm$ 0.05	& (1)\\
M$_\mathrm{WD}$	& 0.63 $\pm$ 0.05 M$_\mathrm{\odot}$ & (1)\\
Distance	& 141.7 $\pm$ 2.5 pc & (2)\\
\enddata
\tablecomments{(1) \citet{Izquierdo2018}; (2) Gaia DR2
}
\end{deluxetable}

The transiting fragments include dust particles and observations at different wavelengths could constrain 
their size and composition -- the main motivation for multi-wavelength photometric observations. Previous 
studies show that the transit depths are the same from V to J band \citep{Alonso2016,Zhou2016,Croll2017}. 
Observations at K$_\mathrm{s}$ band and 4.5~$\mu$m have revealed shallower transits than those in the optical 
\citep{Xu2018a}. However, after correcting for the excess emission from the dust disk, the transit depths 
become the same at all the observed wavelengths. Under the assumption of optically thin transiting material, 
the authors conclude that there is a dearth of dust particles smaller than 1.5~$\mu$m due to their short 
sublimation time \citep{Xu2018a}.

The first detection of a color difference in WD~1145+017's transits is featured by a shallower u'-band transit 
(u' - r' = -0.05 mag) using multi-band fast photometry from ULTRACAM \citep{Hallakoun2017}. The authors have demonstrated that limb darkening cannot reproduce the observed difference in the transit depth and that dust extinction is unlikely to be the mechanism either. Finally, they proposed that the most likely cause is the reduced circumstellar gas absorption during transits because of the high concentration of circumstellar lines in the u' band. Due to the active nature of this system, simultaneous photometric and spectroscopic 
observations are required to test this hypothesis.

In this paper, we report results from a large spectroscopic and photometric campaign of WD~1145+017 with the 
{\it Hubble Space Telescope} ({\it HST}), the {\it Keck Telescope}, the {\it Very Large Telescope} ({\it VLT}), 
{\it Spitzer Space Telescope}, and several smaller telescopes. The main goal is to understand the interplay between the transiting fragments, dust 
disk, and circumstellar gas. This paper is organized as follows. Observations and data reduction of the main 
dataset are presented in Section~\ref{sec:obs_dr}. The light curves are analyzed in 
Section~\ref{sect:uv_transit}, which features shallow 4.5~$\mu$m and UV transits. All the spectroscopic analysis 
is presented in Section~\ref{sect:spectra}, where we found an anti-correlation between the optical transit depth and 
the strength of the circumstellar absorption lines. In Section~\ref{sec:interp}, we present a model that could 
quantitatively explain the shallow UV transits from the change of circumstellar lines. Conclusions are presented 
in Section~\ref{sec:final}.

\section{Observations and Data Reduction \label{sec:obs_dr}}

\subsection{UV photometry and spectroscopy}

\begin{figure}
\includegraphics[width=0.45\textwidth]{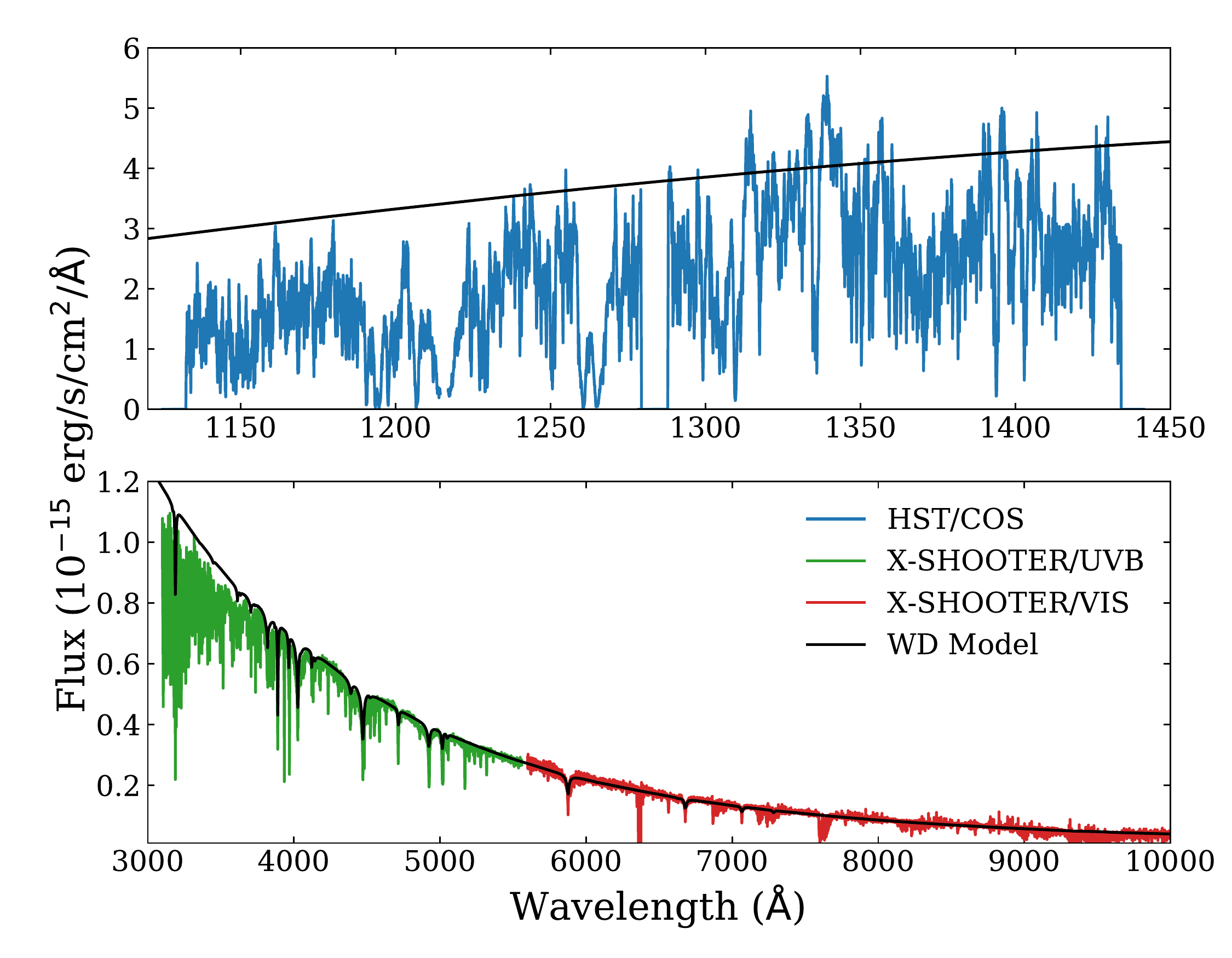}
\caption{A representative UV and optical spectrum of WD~1145+017 from {\it HST}/COS (top panel) and {\it 
VLT}/X-SHOOTER (bottom panel). Also shown is a metal-free helium-dominated white dwarf model (black solid line) 
with parameters appropriate for WD~1145+017 (see Section 4). In the COS wavelength region, there are many 
absorption features and 
essentially no measured continuum.
\label{fig:whole_spectra}}
\end{figure}

We were awarded observing time with the Cosmic Origins Spectrograph (COS) onboard {\it HST} to observe 
WD~1145+017 (program ID \#14467, \#14646, \#15155) a few times between 2016 and 2018. The observing log is 
listed in Table~\ref{tab:cos_log}. The G130M grating was used with a central wavelength of 1291 {\AA} and a 
wavelength coverage of 1125 to 1440 {\AA}. To minimize the effect of fixed pattern noise, we adopted different 
FP-POS steps for each exposure (COS Instrument Handbook). As a result, the wavelength coverage was slightly different in each exposure. A representative 
COS spectrum, full of absorption features, is shown in Fig.~\ref{fig:whole_spectra}.

\begin{figure*}
\gridline{\fig{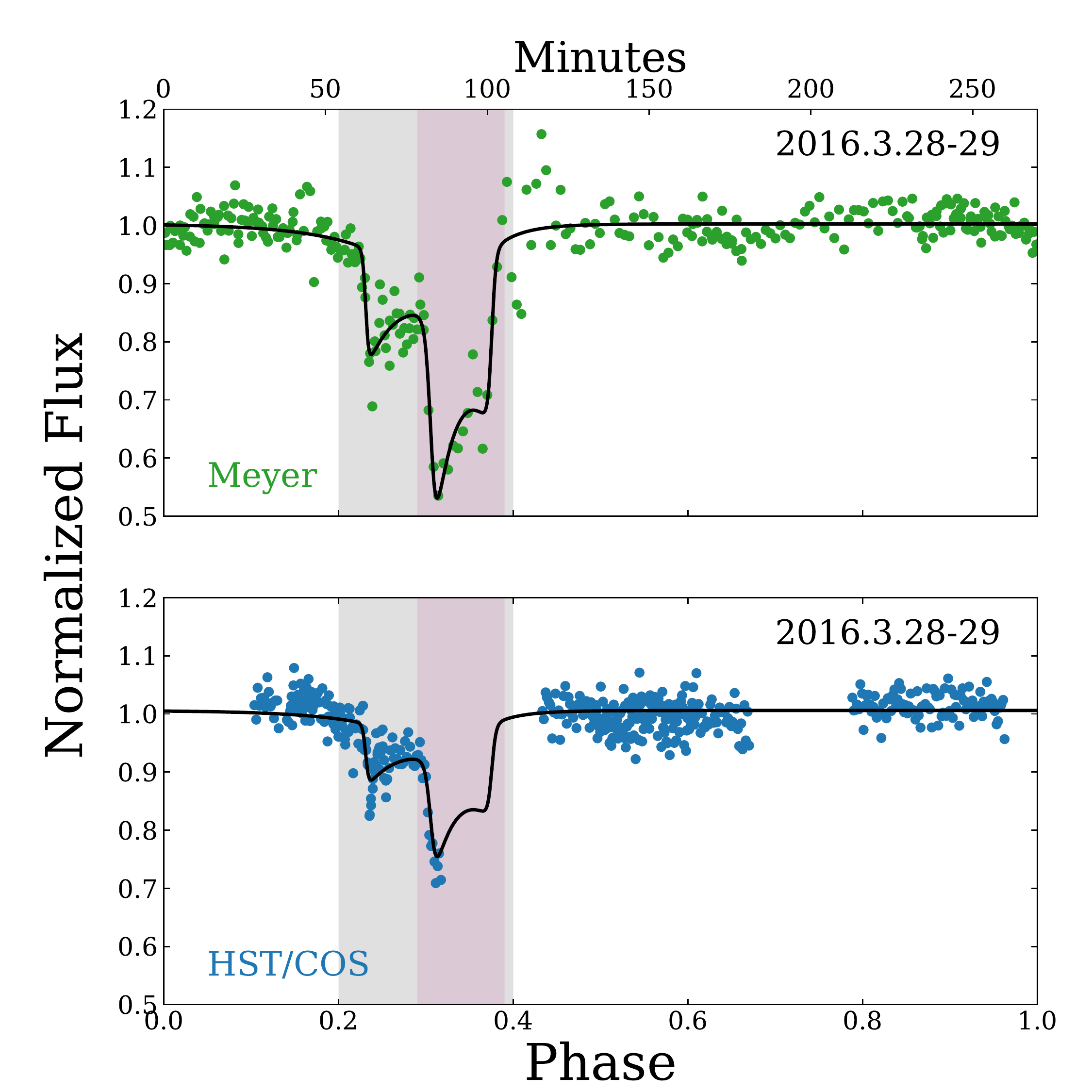}{0.5\textwidth}{}
\fig{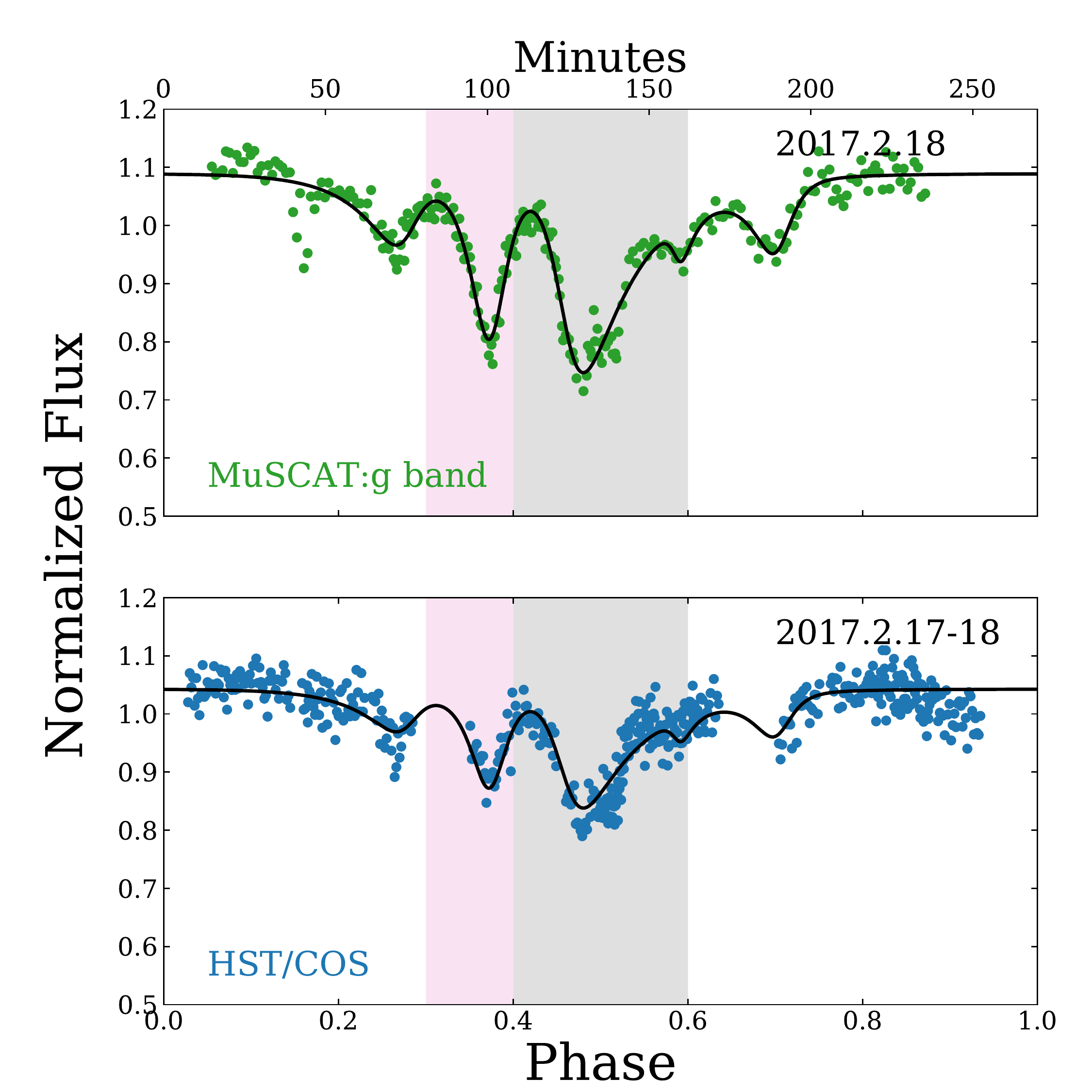}{0.5\textwidth}{} }
\gridline{\fig{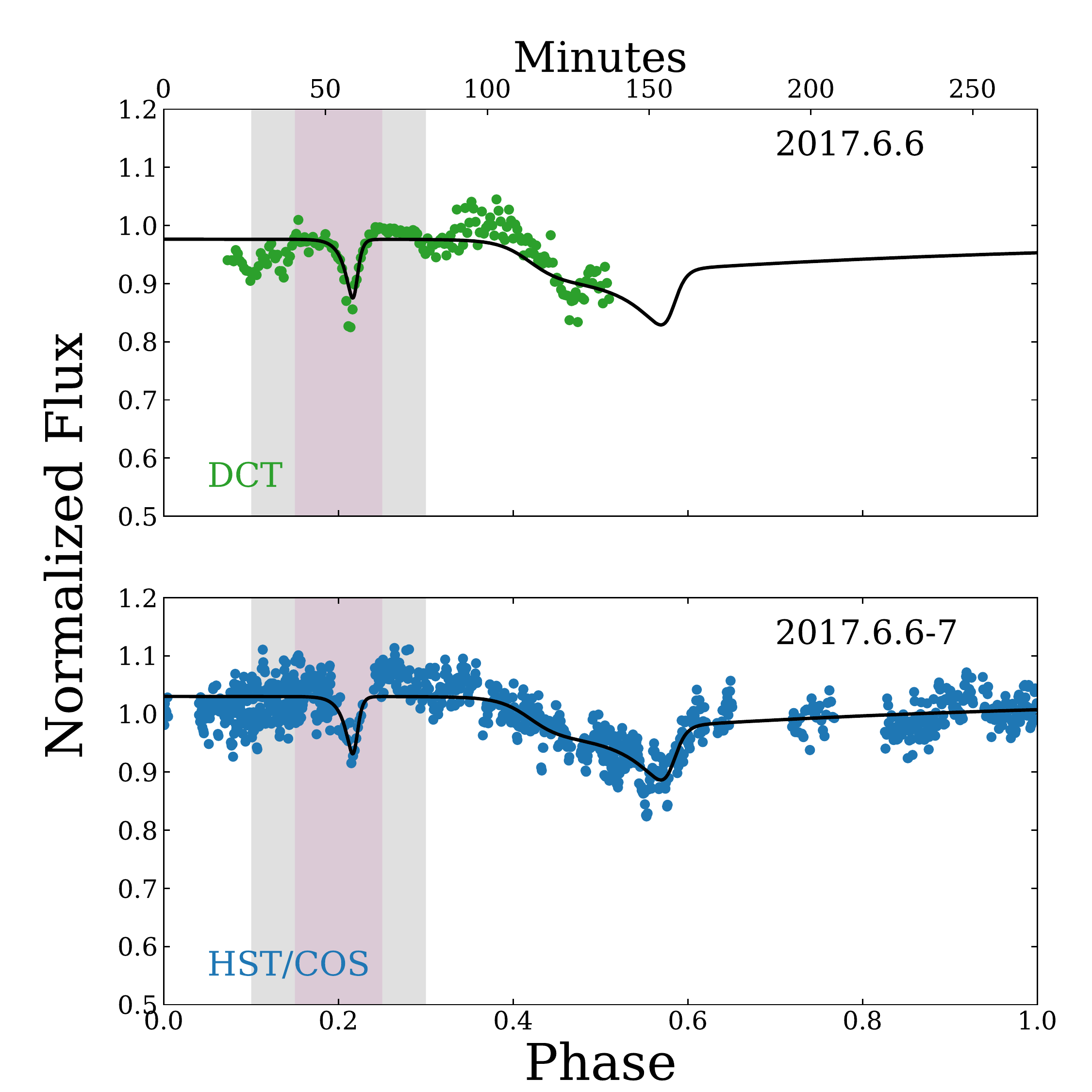}{0.5\textwidth}{}
\fig{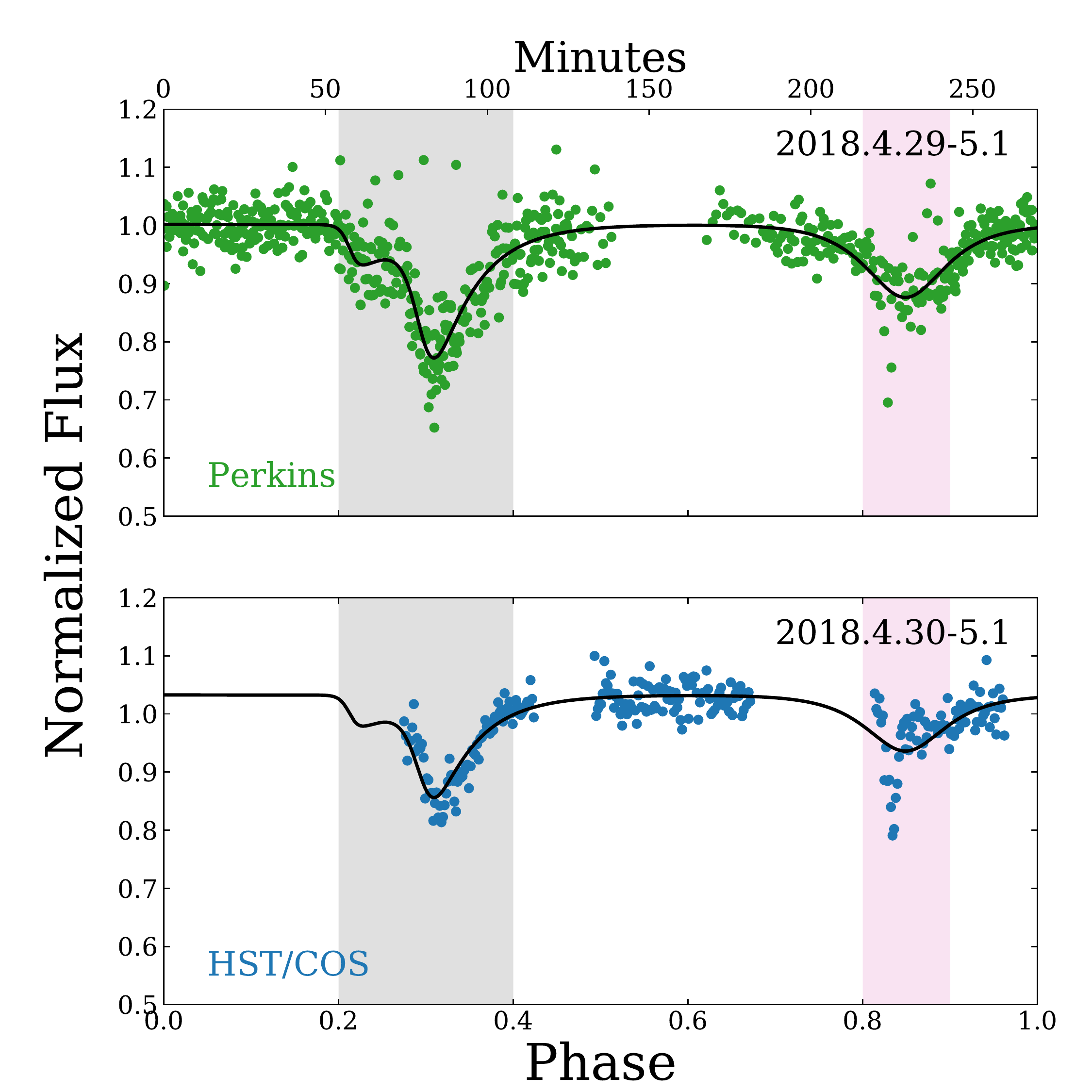}{0.5\textwidth}{} }
\caption{UV and optical light curves of WD~1145+017 from 2016 to 2018. The optical and UV data were taken within one night.  The black line represents the best fit model for a given night. The pink and grey shaded area marks the phase range used to calculate the average transit depth in Table~\ref{tab:transit_par}.
\label{fig:uv_lc}}
\end{figure*}

To extract the light curve from the time-tagged COS observations, we used a \textsc{Python} library created under the Archival Legacy Program (ID: \# 13902, PI: J. Ely, ``The Light curve Legacy of COS and STIS", see also \citealt{Sandhaus2016}). This library begins with the event-list datasets produced as part of a routine \textsc{CalCOS} reduction, and performs additional filtering, calibration, and extraction in order to transform a spectral dataset into the time domain sequence.  This extraction can be done with any time sampling and wavelength ranges. Here, we selected the wavelength range that was shared by all the FP-POS exposures at each epoch and a time sampling of 30~sec. The light curve was normalized by dividing by a constant, which is the average flux of the out-of-transit light curve, as shown in Fig.~\ref{fig:uv_lc}. 

In the 97 minute orbit of {\it HST}, WD~1145+017 is visible for about 50 minutes. The orbital period of the fragment is $\approx$  270~minutes, about a factor of 3 times the {\it HST} orbital period. In 2016, five consecutive {\it HST} orbits were executed, covering three separate parts of the orbital phase. In 2017, we improved our observing strategy by setting up two groups of observations with 3 orbits each, separated by 10 orbits in between. This set-up allows us to have an almost complete phase coverage. In 2018, we used a similar set-up as those in 2017. Unfortunately due to a gyro failure, only 3 orbits worth of data were obtained.

\begin{deluxetable}{lllllll}
\tablecaption{Observing Log of Photometric Observations \label{tab:cos_log}}
\tablewidth{0pt}
\tablehead{
\colhead{Tel./Inst.} & \colhead{$\lambda$ ($\mu$m)}  & \colhead{Date (UT)}
}
\startdata
COS	& 0.13	& Mar 28, 23:15 - Mar 29, 06:19, 2016 \\
Meyer	& 0.48 	& Mar 28, 21:00 - Mar 29, 04:37, 2016 \\
COS	& 0.13	& Feb 17, 11:29 - Feb 18, 07:09, 2017\\
MuSCAT & 0.48	& Feb 18, 13:18 - Feb 18, 20:35, 2017\\
COS & 0.13	& Jun 6, 07:02 - Jun 7, 04:07, 2017 \\ 
DCT &0.55 & Jun 6, 03:35 - Jun 6, 05:32, 2017\\
IRAC & 4.5 & Apr 25, 10:15- Apr 25, 20:05, 2018 \\
NASACam	& 0.55 & Apr 25, 3:51 - Apr 25, 9:19, 2018 \\
NASACam	& 0.55 & Apr 26, 4:58 - Apr 26, 8:21, 2018\\
COS & 0.13	& Apr 30, 22:57 - May 1, 18:42, 2018 \\
Perkins & 0.65 & Apr 29, 03:55 - May 1, 08:10, 2018\\
\enddata
\end{deluxetable}

\subsection{Spitzer Photometry}

\begin{figure}
\includegraphics[width=0.45\textwidth]{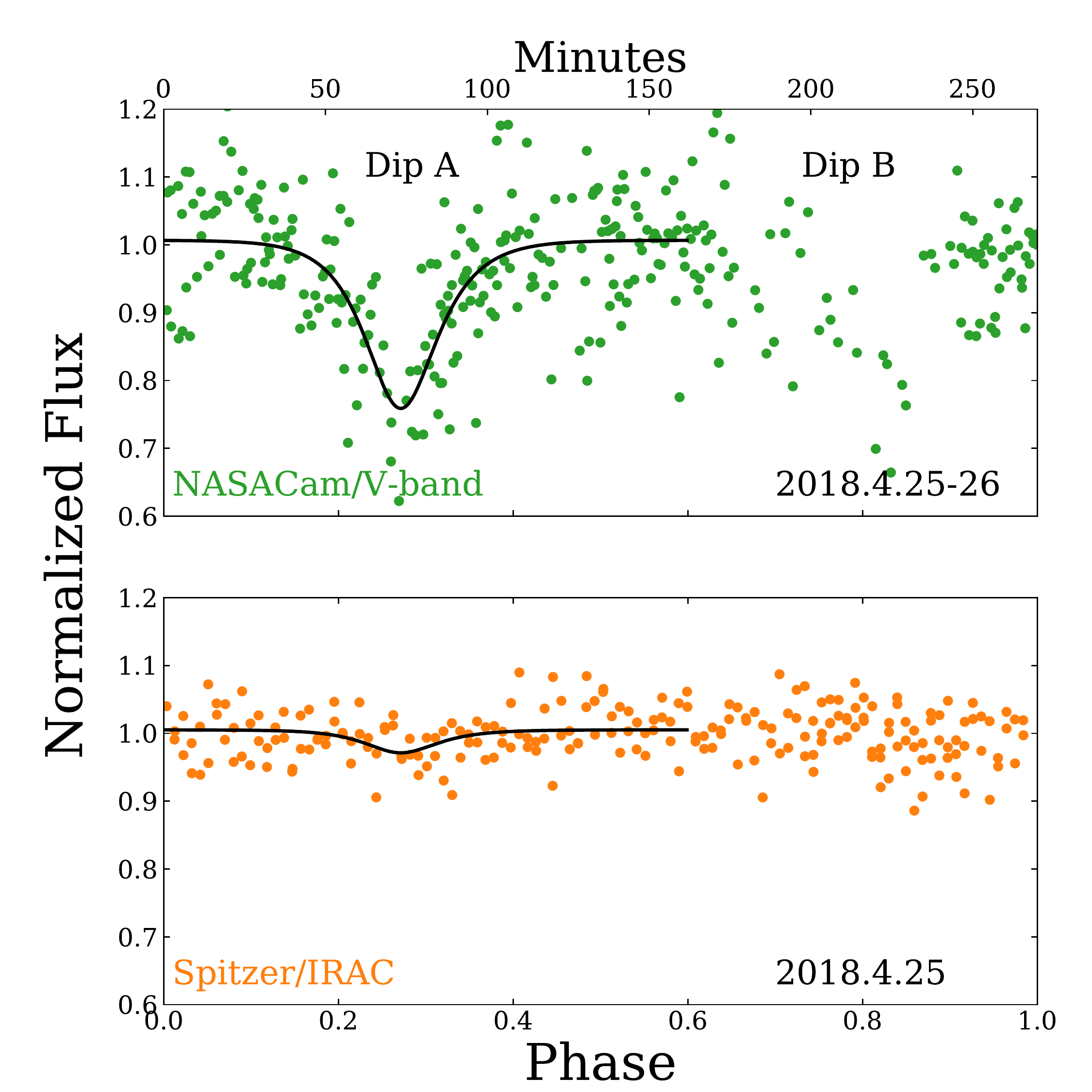}
\caption{Spitzer 4.5~$\mu$m and optical light curves of WD~1145+017 in 2018. The black line represents the best fit model to the data. \label{fig:spitzer_lc}}
\end{figure}

WD~1145+017 was observed with {\it Spitzer}/IRAC at 4.5~$\mu$m under 
program \#13065. Following our previous set-up on the same target \citep{Xu2018a}, the science 
observation had 1140 exposures with 30~sec frame time in stare mode. The total on-target time was 9.5~hr, 
covering a little over two full 4.5-hr cycles. Data reduction was performed following procedures outlined in \citet{Xu2018a}, 
and the light curve is shown in Fig.~\ref{fig:spitzer_lc}. There is one transit (Dip A) marginally detected 
with Spitzer around phase 0.3.

\subsection{Optical Photometry \label{sec:optical_phot}}

We have arranged optical photometric monitoring around the same time of {\it HST} and {\it Spitzer} observations. We 
present the highest quality optical light curves here; the logs are listed in Table~\ref{tab:cos_log}. We 
briefly describe each observation.

{\it 2016 Meyer observation}: The 0.6m Paul and Jane Meyer Observatory Telescope in the Whole Earth Telescope (WET) network \citep{Provencal2012} was used for observing WD~1145+017 during the 2016 {\it HST} window. The exposure time is 60 sec and weather conditions were moderate with some passing clouds. This epoch of observation has been reported in \citet{Xu2018a}, together with simultaneous {\it VLT}/K$_\mathrm{s}$ band and {\it Spitzer} 4.5~$\mu$m observations. 

{\it 2017 MuSCAT observation}: MuSCAT is mounted at the 188-cm telescope at the Okayama Astrophysical Observatory in Japan \citep{Narita2015}. We observed the target in g, r, and z$_\mathrm{s}$ bands simultaneously with 60 sec exposure time in each filter. No color difference was detected and the g-band light curve is presented, which has the highest quality.

{\it 2017 DCT observation}: We observed WD~1145+017 from the 4.3-m Discovery Channel Telescope (DCT) using the Large 
Monolithic Imager (LMI, \citealt{Massey2013}). A series of 30~s exposures in V-band were taken using 2x2 pixel 
binning under thin cirrus. The data reductions were similar to those described in 
\citet{DalbaMuirhead2016,Dalba2017}. Aperture photometry was performed on the target as well as 
suitable reference stars to generate light curves. This procedure was repeated for a range of photometric 
apertures. The aperture and the set of reference stars that maximized the photometric precision away from the 
dimming events were used to generate the final light curve.

{\it 2018 NASACam Observation}: On both Apr 25 and 26, the V-band filter was used with an exposure time of 60~sec. These data were reduced and analyzed following the DCT observations. As shown in Fig.~\ref{fig:spitzer_lc}, the optical light curve was rather noisy due to the proximity ($<$30 deg) of the nearly fully illuminated ($>$75\%) Moon. Nevertheless, Dip A is well detected in the optical light curve and there is also a weaker Dip B around phase 0.8. Our long-term monitoring of WD~1145+017 over this period ({\it Gary et al. in prep}) shows that Dip B is changing quickly, while Dip A is relatively stable. We will focus on Dip A for the following analysis.

{\it 2018 Perkins Observation}: The PRISM instrument \citep{Janes2004} mounted on the 1.8-m Perkins telescope at Lowell Observatory was used to observe WD~1145+017. The exposure times were 45~s. On April 29, the sky was clear but it was windy. The wind continued on April 30 and patchy cirrus was present. On both nights, the seeing was consistently greater than 2{\farcs}0. These observations were reduced and analyzed in the same fashion as the DCT observations.

In addition, WD 1145+017 has been monitored on a regular basis by amateur astronomers using the 14-inch telescope at the Hereford Arizona Observatory (HAO) and a 32-inch telescope at Arizona (some of the observations have been reported in \citealt{Rappaport2016, Alonso2016, Rappaport2017, Gary2017}; see details in those references). We also utilized light curves obtained from the University of Arizona's 61 inch telescope, a 14-inch telescope at Cyprus, a 32-inch IAC80 telescope on the Canary Islands, as well as a 20-inch telescope in Chile. We use optical photometric observations that were taken closest in time with our spectroscopic monitoring described in the next section. 

\subsection{Optical Spectroscopy \label{sec:spectroscopy}}

WD~1145+017 has been observed intensively with different optical spectrographs from 2015-2018. A summary of the observing log is listed in Table~\ref{tab:spectra_log}. 

{\it Keck}/HIRES: High Resolution Echelle Spectrometer (HIRES; \citealt{Vogt1994}) on the {\it Keck} I 
telescope has a blue collimator and a red collimator. HIRESb was used more frequently for observing WD~1145+017 
because it covers shorter wavelengths, where most circumstellar lines are located, approximately the green 
region (X-SHOOTER UVB arm) shown in the lower panel of
Fig.~\ref{fig:whole_spectra}. For HIRES observations, the C5 decker was used with a slit width of 
1{\farcs}148, returning a spectral resolution of 40,000. The exposure times vary to ensure a signal-to-noise 
ratio of at least 10 in a single exposure. Data reduction was performed with the MAKEE package. We also 
continuum normalized the spectra with IRAF following procedures described in an early study of WD~1145+017 
\citep{Xu2016}. Thanks to the high spectral resolution of HIRES, we can resolve the profiles of individual 
absorption lines and separate the photospheric component from the circumstellar component.

{\it Keck}/ESI: WD~1145+017 has also been observed with the Echellette Spectrograph and Imager (ESI; 
\citealt{Sheinis2002}) on the {\it Keck} II telescope. A slit width of 0{\farcs}3 was used, which returns a 
spectral resolution of 14,000. Similar to the HIRES analysis, data reduction was performed using both MAKEE and 
IRAF. The main advantage of ESI is its wider wavelength coverage and shorter integration times due to its lower 
spectral resolution. The ESI dataset is well suited to probe short-term (tens of minutes) variations of the 
circumstellar lines.

{\it VLT}/X-SHOOTER: WD~1145+017 has been observed with X-SHOOTER \citep{Vernet2011} on the {\it VLT} at the 
same time as the 2016 {\it HST} observation. The weather conditions were decent with some thin clouds. 
X-SHOOTER has three arms, i.e. UVB, VIS, and NIR, which provide simultaneous wavelength coverage from the 
atmospheric cutoff in the blue to K band. WD~1145+017 is too faint to be detected in the NIR arm 
and here we focus on the UVB and VIS arms. A series of short exposures was taken to monitor the variations of 
the circumstellar lines. The data were reduced using the XSHOOTER pipeline 2.7.0b by the ESO quality control 
group. The final combined spectrum is presented in Fig.~\ref{fig:whole_spectra}.

\begin{deluxetable*}{lllllll}
\tablecaption{Optical Spectroscopic Observations of WD~1145+017 \label{tab:spectra_log}}
\tablewidth{0pt}
\tablehead{
\colhead{Instrument} & \colhead{Wavelength} & \colhead{Resolution}   & \colhead{Date (UT)}& \colhead{Exposure Times}
}
\startdata
{\it Keck}/HIRESb	& 3200--5750 {\AA} & 40,000 & 2015 Apr 11	& 2400s$\times$3\\
{\it Keck}/ESI	& 4700--9000 {\AA} & 14,000	& 2015 Apr 25	& 1180s $\times$ 2\\
{\it Keck}/HIRESr	& 4700--9000 {\AA} & 40,000	& 2016 Feb 3	& 1800s$\times$9\\
{\it Keck}/HIRESb	&  3200--5750 {\AA} & 40,000	& 2016 Mar 3	& 1259s, 2400s$\times$5, 2000s, 1440s, 2400s$\times$2\\
{\it Keck}/ESI		& 3900--9300 {\AA} & 14,000	& 2016 Mar 28	& 1300s, 900s$\times$2, 600s$\times$25\\
{\it VLT}/X-SHOOTER&	3100--10,000 {\AA}&	6200/7400\tablenotemark{a}& 2016 Mar 29	& 280s/314s\tablenotemark{a} $\times$ 29\\
{\it Keck}/HIRESb	& 3100--5950 {\AA}& 40,000	& 2016 Apr 1	& 1200s$\times$9\\
{\it Keck}/ESI		& 3900--9300 {\AA} & 14,000	& 2016 Nov 18	& 600s$\times$6\\
{\it Keck}/ESI		& 3900--9300 {\AA} & 14,000	& 2016 Nov 19	& 600s$\times$11\\
{\it Keck}/HIRESr	& 4715--9000 {\AA}	& 40,000 & 2016 Nov 26	& 1800s $\times$2 \\
{\it Keck}/HIRESr	& 4780--9200 {\AA}	& 40,000	& 2016 Dec 22	& 1500s \\
{\it Keck}/ESI		& 3900--9300 {\AA} & 14,000	& 2017 Mar 6	& 600s$\times$3, 480s$\times$3\\
{\it Keck}/ESI		& 3900--9300 {\AA} & 14,000	& 2017 Mar 7	& 600s$\times$6\\
{\it Keck}/ESI		& 3900--9300 {\AA} & 14,000	& 2017 Apr 17	& 500s$\times$8\\
{\it Keck}/HIRESb	& 3100--5950 {\AA}& 40,000	& 2018 Jan 1	& 900s$\times$5 \\
{\it Keck}/HIRESb	& 3100--5950 {\AA}& 40,000	& 2018 Apr 24	& 1200s$\times$5, 1000s, 1350s$\times$ 2 \\
{\it Keck}/HIRESb	& 3100--5950 {\AA}& 40,000	& 2018 May 18	& 1200$\times$5 \\
\enddata
\tablenotetext{a}{The first number is for the UVB arm while the second number is for the VIS arm.}
\end{deluxetable*}

\section{Transit Analysis \label{sect:uv_transit}}

To model the light curve, we follow previous studies of fitting asymmetric transit \citep[e.g.][]{Rappaport2014} by adopting a series of asymmetric hyperbolic secant (AHS) functions with the following form:
\begin{equation}
f(p)=f_{\mathrm{0}} \, \left[1 - f_{\mathrm{dip}}(p)\right] ~{\equiv}~ f_\mathrm{0} \, \left(1 - \sum \limits_{i}  \frac{2f_\mathrm{i}}{e^\frac{p-p_i}{\phi_{i1}}+e^{-\frac{p-pi}{\phi_{i2}}}} \right) . 
\label{Equ:AHS}
\end{equation}  
where $i$ represents the number of AHS component, $p_i$ is the phase of the deepest point in a transit, $\phi_{i1}$ and $\phi_{i2}$ represent the phase duration of the ingress and egress, respectively.

Assuming the light curves have the same shape but different depth at different wavelengths, we can fit the light curve at another wavelength as:
\begin{equation}
f_\mathrm{\lambda}(p)=f_\mathrm{\lambda,0} \, \left[1 - d_\mathrm{\lambda} \, f_\mathrm{dip}(p) \right]   
\label{Equ:d}
\end{equation}
There are only two free parameters, $f_\mathrm{\lambda,0}$, which characterizes the continuum level, and $d_\mathrm{\lambda}$, which represents the transit depth ratios. $f_\mathrm{dip}(p)$ can be taken from the best fit parameters from Equ.~(\ref{Equ:AHS}).

\subsection{4.5$\mu$m and Optical Transits \label{sec:spitzer}}

We fitted Dip A with one AHS function and the best fit models are shown as black lines in Fig.~\ref{fig:spitzer_lc}. We calculated the transit depth ratios between 4.5~$\mu$m and optical, as listed in Table~\ref{tab:spitzer_D}. We also list the ratios calculated for 2017 and 2018 epochs \citep{Xu2018a} and take the average value d$_\mathrm{4.5\mu m}$/d$_\mathrm{opt}$ of 0.235 $\pm$ 0.024. After correcting for the contribution from the dust disk at 4.5~$\mu$m, the 
transit depth ratio between 4.5~$\mu$m and optical is 0.995 $\pm$ 0.119, consistent with unity. 

\begin{deluxetable}{llllllll}[ht!]
\tablecaption{Spitzer and Optical Transit Measurements \label{tab:spitzer_D}}
\tablewidth{0pt}
\tablehead{ \colhead{Epoch} & \colhead{Dip} &  \colhead{d$_\mathrm{4.5\mu m}$/d$_\mathrm{opt}$}   }
\startdata
2016	& A1	& 0.256 $\pm$ 0.032 \\
2017	& B2 & 0.309 $\pm$ 0.075 \\
2017	& B3	& 0.239 $\pm$ 0.026 \\
2018	& A	& 0.137 $\pm$ 0.044\\
Average	&	& 0.235 $\pm$ 0.024 \\
\enddata
\end{deluxetable}

\begin{figure} 
\includegraphics[width=0.45\textwidth]{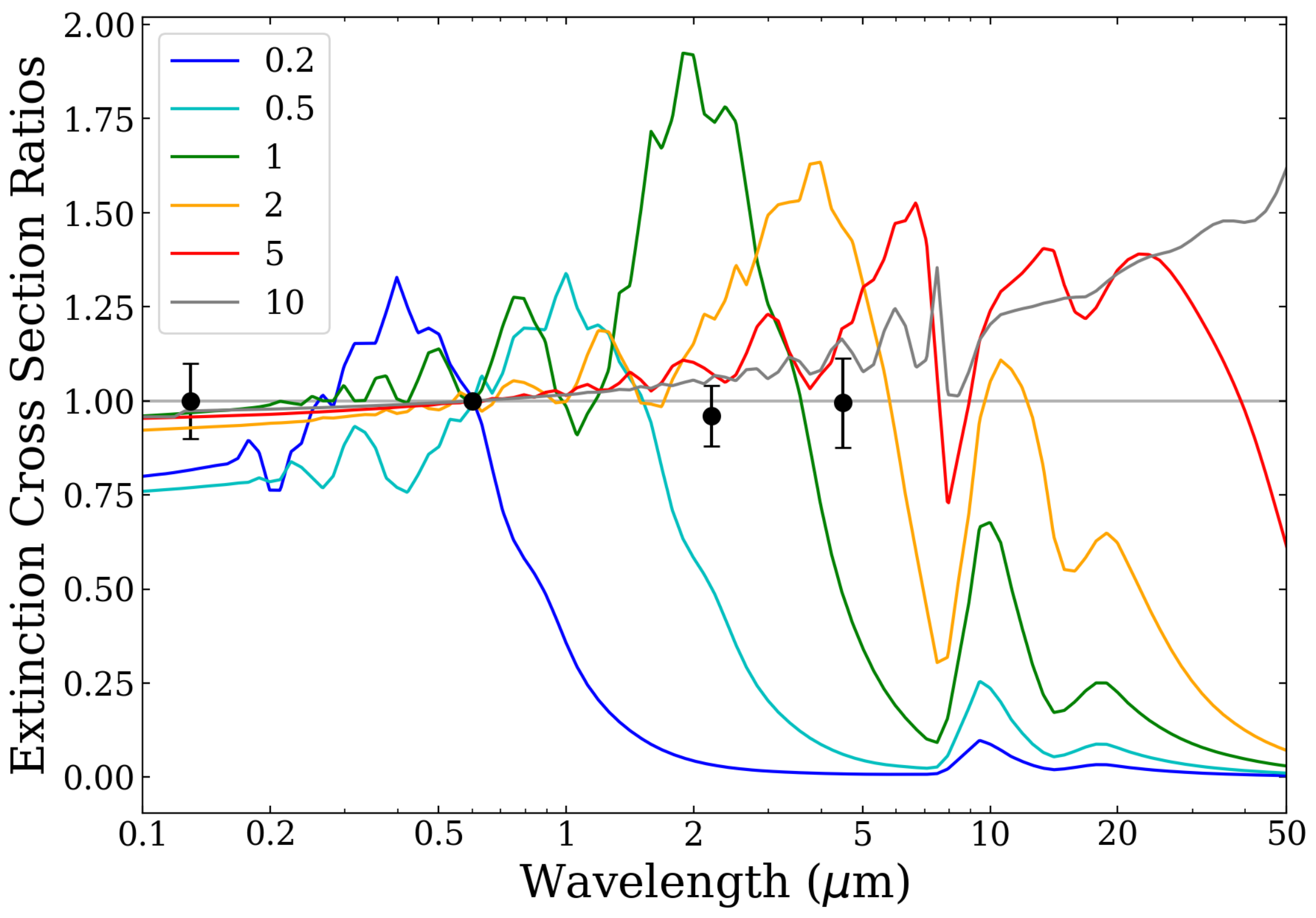}
\caption{Mie extinction cross-section ratios versus wavelength using astronomical silicates \citep{DraineLee1984, LaorDraine1993} for a particle radius $s$ from 0.2 to 10~$\mu$m. For each curve, a bulge peaks at $s$ $\times$ $\pi$/2 is a characteristic of the Mie scattering cross section. The other two peaks at 10 $\mu$m and 20 $\mu$m are the absorption cross section of astronomical silicates. Black dots are the measured transit depths ratios in WD~1145+017 and grains smaller than 2~$\mu$m are inconsistent with the observation
\label{fig:mie}
}
\end{figure}

In Fig.~\ref{fig:mie}, we compared the new measurements with Mie scattering cross sections of astronomical silicates calculated by \citet{DraineLee1984} and \citet{LaorDraine1993}. The astronomical silicates is not a real mineral, and its real and imaginary refractive index are calculated from a combination of lab measurements and actual astronomical observation of circumstellar and interstellar silicate dust \citep{DraineLee1984}. Our transit observations cover 0.12~$\mu$m\footnote{The 0.12~$\mu$m COS light curve is analyzed in Section~\ref{sec:interp}.} to 4.5~$\mu$m and yet no wavelength dependence caused by the transiting material has been detected. The result is consistent with our previous finding that either the transiting material has very few small grains or it is optically thick \citep{Xu2018a}. 

In addition, the out-of-transit flux at 4.5~$\mu$m in 2018 is 52.7 $\pm$ 4.7~$\mu$Jy, which is consistent 
with 55.0 $\pm$ 3.2~$\mu$Jy reported in 2016 and 2017 \citep{Xu2018a}. The infrared fluxes of 
WD~1145+017 are surprisingly stable given the transit behavior has changed dramatically during the past few 
years. A recent study by \citet{Swan2019} has found that WD~1145+017 is variable in the {\it WISE} W1 band 
(3.4~$\mu$m). The cause is unclear due to the sparse sampling of {\it WISE} and the variability could either come from the transits or the disk material. Infrared 
variability of white dwarf dust disks has been reported up to 30\% at the IRAC bands \citep{XuJura2014,Xu2018b,Farihi2018c} 
and the proposed scenario is tidal disruption and dust production/destruction. It is evident that dust is 
constantly produced and destroyed around WD~1145+017 yet its infrared fluxes remained unchanged. In addition, the canonical geometrically thin optically thick disk aligned with the transiting material can not reproduce the strong infrared excess \citep{Xu2016}. Either the dust disk is misaligned with the transiting object or the disk has a significant scale height. Numerical simulations have shown that white dwarf dust disks can have a vertical structure when there is a constant input of material \citep{KenyonBromley2017a,KenyonBromley2017b}. However, it is still a puzzle given the stable 4.5~$\mu$m flux.

\subsection{UV and Optical Transits}

For the 2016, February 2017, and 2018 observations, we started by fitting the optical light curves with AHS functions because they have good phase coverage. For the June 2017 observations, we started with the UV light curve because the optical light curve only covers a small phase range. The best fit models are shown in Fig.~\ref{fig:uv_lc} and the UV-to-optical transit depth ratios are listed in Table~\ref{tab:transit_par}. Here, we are reporting the average ratio for a given epoch and this number could differ for each dip. The UV transit depths are always shallower than those of the optical, which is rather surprising given the constant 
transit depth from optical to 4.5~$\mu$m presented in Sectio~\ref{sec:spitzer}. In addition, the UV-to-optical transit ratios also appear to be changing at different epochs.

To quantify the strength of a transit, we introduce $\overline{D}$,
\begin{equation} \label{equ:D}
\overline{D}= \frac{1}{p_\mathrm{2} - p_\mathrm{1}} \times \int_\mathrm{p_\mathrm{1}}^{p_\mathrm{2}} f_\mathrm{0} \times f_\mathrm{dip}(p) dp,
\end{equation}
where p1 and p2 represent the phase interval of interest and $f_\mathrm{0}$ and f$_\mathrm{dip}$(p) are taken from the Equ~(\ref{Equ:AHS}). It is similar to the mean transit depth used in \citet{Hallakoun2017}.

For each epoch, we selected two phase ranges and calculated the corresponding transit depths, as listed in Table~\ref{tab:transit_par}. The transit depths are different for different dips.

\begin{deluxetable*}{lcccccc}[ht!]
\tablecaption{UV and Optical Transit Measurements\label{tab:transit_par}}
\tablewidth{0pt}
\tablehead{
\colhead{Date} & \colhead{$\overline{d}_\mathrm{uv}$/$\overline{d}_\mathrm{opt}$\tablenotemark{a}} & \colhead{Phase}& \colhead{$\overline{D}_\mathrm{opt}$\tablenotemark{b}}   & \colhead{$\bar{\alpha}$\tablenotemark{c}} & \colhead{$\bar{\beta}$\tablenotemark{c}}
}
\startdata
2016 Mar 28-29 &0.531 $\pm$ 0.020 & 0.29-0.39	& 0.295 $\pm$ 0.038 & 0.705 $\pm$ 0.038 & 0.373 $\pm$ 0.103\\
 & & 0.20-0.40		& 0.209 $\pm$ 0.026 & 0.791 $\pm$ 0.026 & 0.556 $\pm$ 0.071\\
2017 Feb 18-19& 0.625 $\pm$ 0.018  &0.30-0.40	&0.150 $\pm$ 0.015 & 0.850 $\pm$ 0.015 & 0.715 $\pm$ 0.038\\
&&0.40-0.60	 &0.188 $\pm$ 0.016 & 0.812 $\pm$ 0.016 & 0.642 $\pm$ 0.045\\
2017 Jun 6-7  & 0.924 $\pm$ 0.068 & 0.15-0.25		&0.021 $\pm$ 0.006 & 0.979 $\pm$ 0.006 & 0.975 $\pm$ 0.009\\
&&  0.10-0.30	& 0.011 $\pm$ 0.003 & 0.989 $\pm$ 0.003 & 0.987 $\pm$ 0.004\\
2018 Apr 30-May 1& 0.746 $\pm$ 0.036& 0.80-0.90 &0.102 $\pm$ 0.009 & 0.898 $\pm$ 0.009 & 0.835 $\pm$ 0.020\\
 & & 0.20-0.40 & 0.107 $\pm$ 0.030 & 0.893 $\pm$ 0.030 & 0.828 $\pm$ 0.051\\
\enddata
\tablenotetext{a}{The average optical-to-UV transit depth ratio for a given epoch.}
\tablenotetext{b}{$\overline{D}_\mathrm{opt}$ is the average optical transit depth for the given phase range, as defined in Equ.~(\ref{equ:D}).}
\tablenotetext{c}{$\bar{\alpha}$ and $\bar{\beta}$ are defined in section~\ref{sec:interp}.}
\end{deluxetable*}

\section{Spectroscopic Analysis \label{sect:spectra}}

With this extensive spectroscopic dataset, we updated the white dwarf models to consistently fit the optical and UV spectra of WD~1145+017. We have also developed new models for the circumstellar lines. Details will be presented in {\it Fortin-Archambault et al. (in prep)} and {\it Steele et al (in prep)}.
In the following analysis, we present some preliminary results of the calculation to help 
us understand different components of the absorption features. 

\subsection{Long-term Variability \label{sec:long-term}}

\begin{figure}[ht]
\includegraphics[width=0.48\textwidth]{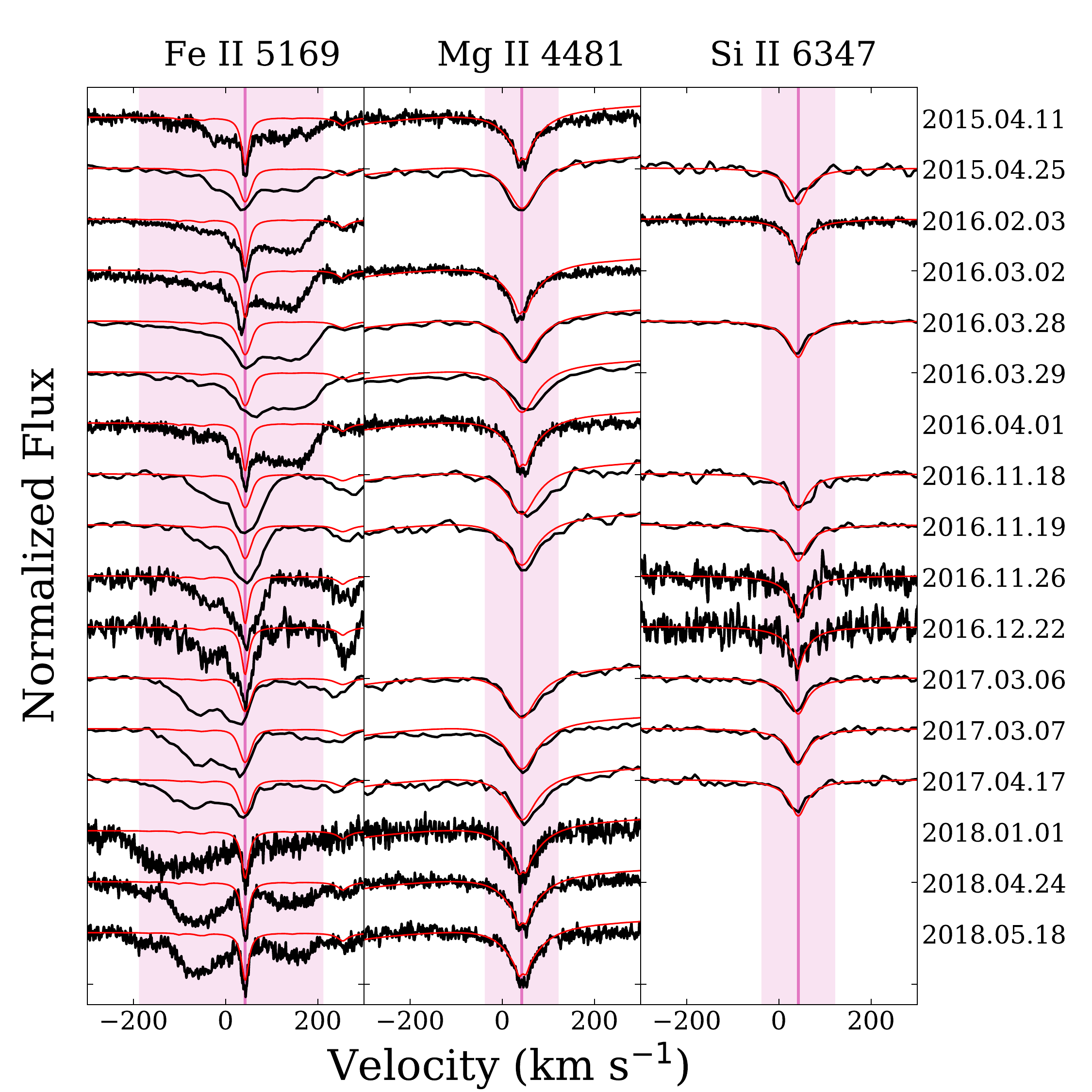}
\caption{A compilation of spectra around Fe II 5169~{\AA}, Mg II doublet 4481~{\AA}, and Si II 6347~{\AA} in the reference frame of the observer. The red line represents our preliminary white dwarf photospheric model spectrum computed to match the instrument resolution (Fortin-Archambault et al. {\it in prep}). The pink line marks the average radial velocity of photospheric lines, which is at~42 km~s$^{-1}$. The pink shaded area marks the wavelength range for calculating the absorptione line strength $\overline{F}_\mathrm{abs}$ defined in Equ.~(\ref{equ:Fabs}). The absorption feature around 250~km~s$^{-1}$ at the Fe II 5169 panel is caused by circumstellar absorption of Mg I at 5173~{\AA}, which is visible during some epochs.
\label{fig:spectra_series}}
\end{figure}

To assess the long-term variability of the absorption features, we selected three representative regions, i.e. Mg II doublet around 4481~{\AA}, Si II 6347~{\AA}, and Fe II 5169 {\AA}, whose transitions come from a lower energy level of  8.86~eV, 8.12~eV, and 2.89~eV, respectively. Both Mg II 4481~{\AA} and Si II 6347~{\AA} primarily have a photospheric contribution due to the high lower energy level of the transition, while Fe II 5169~{\AA} has both photospheric and circumstellar contributions \citep{Xu2016}. The spectra covering those regions are shown in Fig.~\ref{fig:spectra_series}. From 2015 to 2018, the shapes of the circumstellar lines have changed significantly, from being both blue-shifted and red-shifted (April 2015), to mostly red-shifted (March 2016), to mostly blue-shifted (March 2017), and back to being both blue-shifted and red-shifted (May 2018). Likely, our observation has covered a whole precession period of $\sim$~3 years.

In stellar spectroscopy, equivalent width is often used to characterize the strength of an absorption line. However, it is not suitable here because of the irregular line shape. We quantify the average strength of an absorption line over the wavelength range $\Delta \lambda$ as 

\begin{equation} \label{equ:Fabs}
\overline{F}_\mathrm{abs} = 1 - \frac{\sum\limits_{i} F_\mathrm{abs,i} \times d\lambda_\mathrm{i}}{F_\mathrm{WD} \times \Delta \lambda},
\end{equation}
where $\sum\limits_{i} d\lambda_\mathrm{i} = \Delta \lambda$, which is the total absorbing wavelength/velocity range. F$_\mathrm{WD}$ is the white dwarf continuum flux without any absorption from heavy elements, which is approximately 1 for a normalized spectrum. F$_\mathrm{abs,i}$ is the flux of an absorption feature at each wavelength $\lambda_\mathrm{i}$ and $d\lambda_\mathrm{i}$ is the wavelength interval. 

\begin{figure}[ht]
\includegraphics[width=0.48\textwidth]{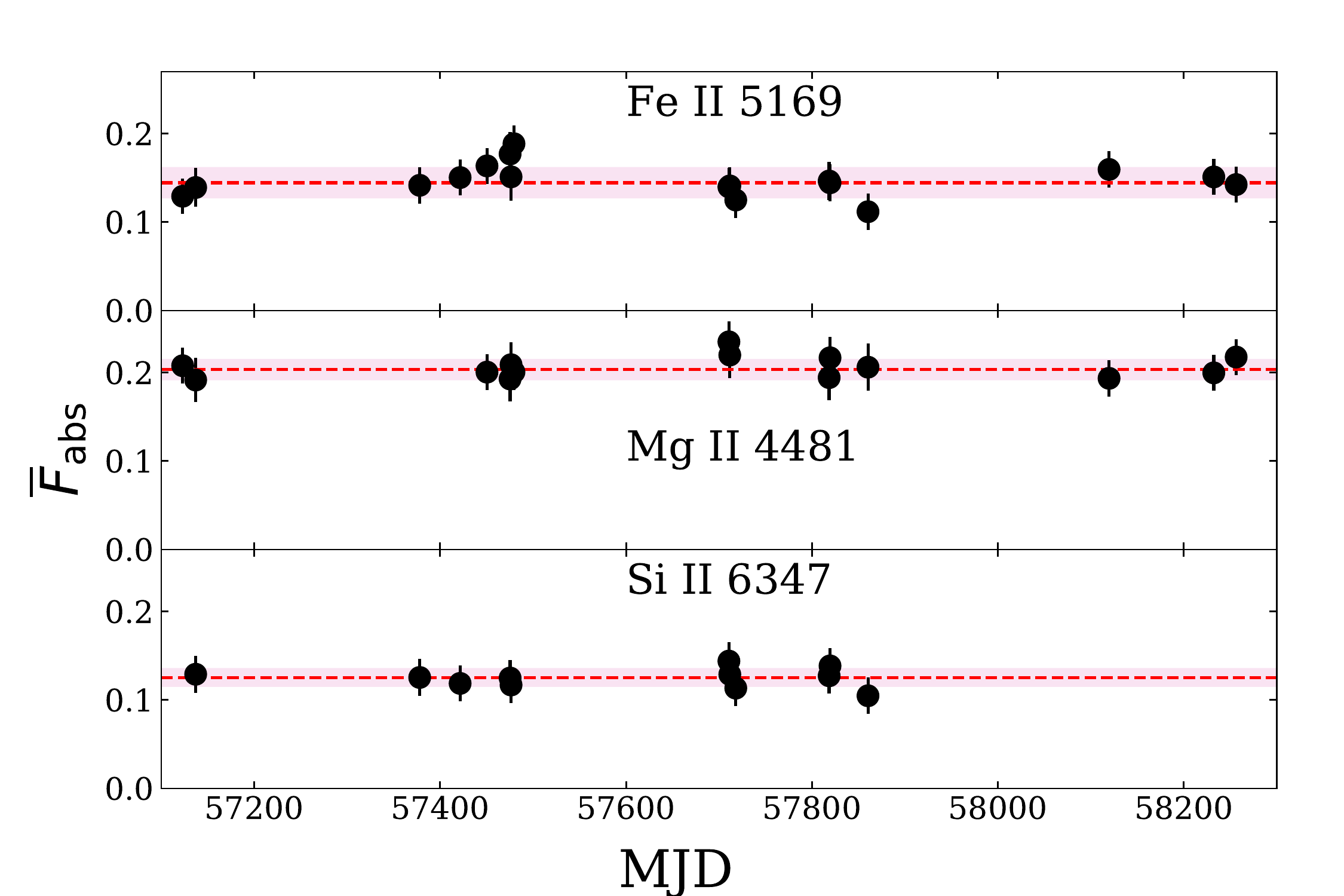}
\caption{Average line strength on a given night as a function of the observing date. The red dashed line indicates the average value and the 
pink shaded area marks the standard deviation. The strength of the absorption lines has 
been the same 
since 2015.}

\label{fig:spectra_time}
\end{figure}

The pink shaded area in Fig.~\ref{fig:spectra_series} marks the velocity range for calculation. For Mg II and Si II, $\overline{F}_\mathrm{abs}$ = $\overline{F}_\mathrm{phot}$, which is the average flux of photospheric absorption because there is little contribution from circumstellar absorption. For Fe II, $\overline{F}_\mathrm{abs}$ = $\overline{F}_\mathrm{phot}$ + $\overline{F}_\mathrm{cs}$, because both photospheric and circumstellar material contribute to the absorption feature. $\overline{F}_\mathrm{cs}$ represents the average flux of circumstellar absorption over a given wavelength range. The average absorption line strength as a function of observing date is shown in Fig.~\ref{fig:spectra_time}. Even though the shapes of the circumstellar lines have changed significantly, the average line strength remains the same from 2015 to 2018, both for circumstellar and photospheric components. Therefore, there are no changes in the compositions of the white dwarf photosphere or the circumstellar material.

\subsection{Line Strength Comparison \label{sec:line_strength}}

The number density of absorption lines is higher in the UV/blue compared to the optical (see Fig.~\ref{fig:whole_spectra} for an example). We selected three regions to assess the overall contribution from photospheric and circumstellar absorption, A. 1330--1420 {\AA} (COS segment A), B. 3200--3900 {\AA} (similar to ULTRACAM u' band reported in \citealt{Hallakoun2017}), C. 6200--6900 {\AA} (similar to ULTRACAM r' band). We can compare the flux of WD~1145+017 in those wavelength ranges with a metal-free white dwarf with the same system parameters (F$_\mathrm{WD}$, which is defined as 100\%) and a polluted white dwarf with only photospheric absorption (for $\overline{F}_\mathrm{phot}$, models are from {\it Fortin-Archambault et al. in prep}). The results are listed in Table~\ref{tab:strength_abs}. 

In the UV, the photospheric and circumstellar lines are ubiquitous, and absorb 43\% of the white dwarf flux. 
In comparison, in the optical, the 
absorbed flux is relatively small compared to the white dwarf flux and the effect becomes even smaller at 
longer wavelengths. As a result, circumstellar lines will have a much larger effect on the overall UV flux than 
the broad-band optical flux, as has been first discussed in \citet{Hallakoun2017}.

We caution that this kind of calculation is very sensitive to the choice of the continuum flux, particularly for the UV observations where there is essentially no measured continuum. We estimated the uncertainty, using the relatively clean part of the spectrum, to be about 5\% for COS observations and 2\% for HIRES observations. That is the dominant source of error in Table~\ref{tab:strength_abs}.

\begin{deluxetable}{lccccc}[ht!]
\tablecaption{WD~1145+017 Flux Comparison \label{tab:strength_abs}}
\tablewidth{0pt}
\tablehead{
\colhead{} & \colhead{2016 Mar 29} & \colhead{2016 Apr 01} & \colhead{2016 Feb 3} \\
\colhead{} & \colhead{COS} & \colhead{HIRESb} & \colhead{HIRESr} \\
\colhead{} & \colhead{(1300--1420 {\AA})}  & \colhead{(3200--3800 {\AA})}   & \colhead{(6100--6700 {\AA})} 
}
\startdata
$F_\mathrm{WD}$& 100 $\pm$ 5\% & 100 $\pm$  2\% & 100 $\pm$ 2\%\\ 
$\overline{F}_\mathrm{abs}$ & 42.7 $\pm$ 5.5\% & 4.2 $\pm$ 2.7\% & 1.3 $\pm$ 2.8\%\\
$\overline{F}_\mathrm{phot}$ & 19.0 $\pm$ 5.4\%& 1.5 $\pm$ 2.8\% & 0.7 $\pm$ 2.8\%\\
$\overline{F}_\mathrm{cs}$ &  23.8 $\pm$ 2.3\% & 2.7 $\pm$ 2.7\% & 0.6 $\pm$ 2.8\%\\
\enddata
\tablecomments{$F_\mathrm{WD}$ is the flux of a metal-free white dwarf with the same parameters as WD~1145+017 (e.g. black line in Fig.~\ref{fig:whole_spectra}). $\overline{F}_\mathrm{abs}$ is calculated directly from the observed spectra, which represents the total amount of photospheric and circumstellar absorption. $\overline{F}_\mathrm{phot}$ is absorption from photospheric absorption, which is calculated from white dwarf models with the same parameters and abundances as WD~1145+017 ({\it Fortin-Archambault et al. in prep}). $\overline{F}_\mathrm{cs}$ is calculated as $\overline{F}_\mathrm{abs}$ - $\overline{F}_\mathrm{phot}$.
}
\end{deluxetable}

\subsection{Short-term Variability \label{sec:short-term}}

\begin{figure}
\includegraphics[width=0.45\textwidth]{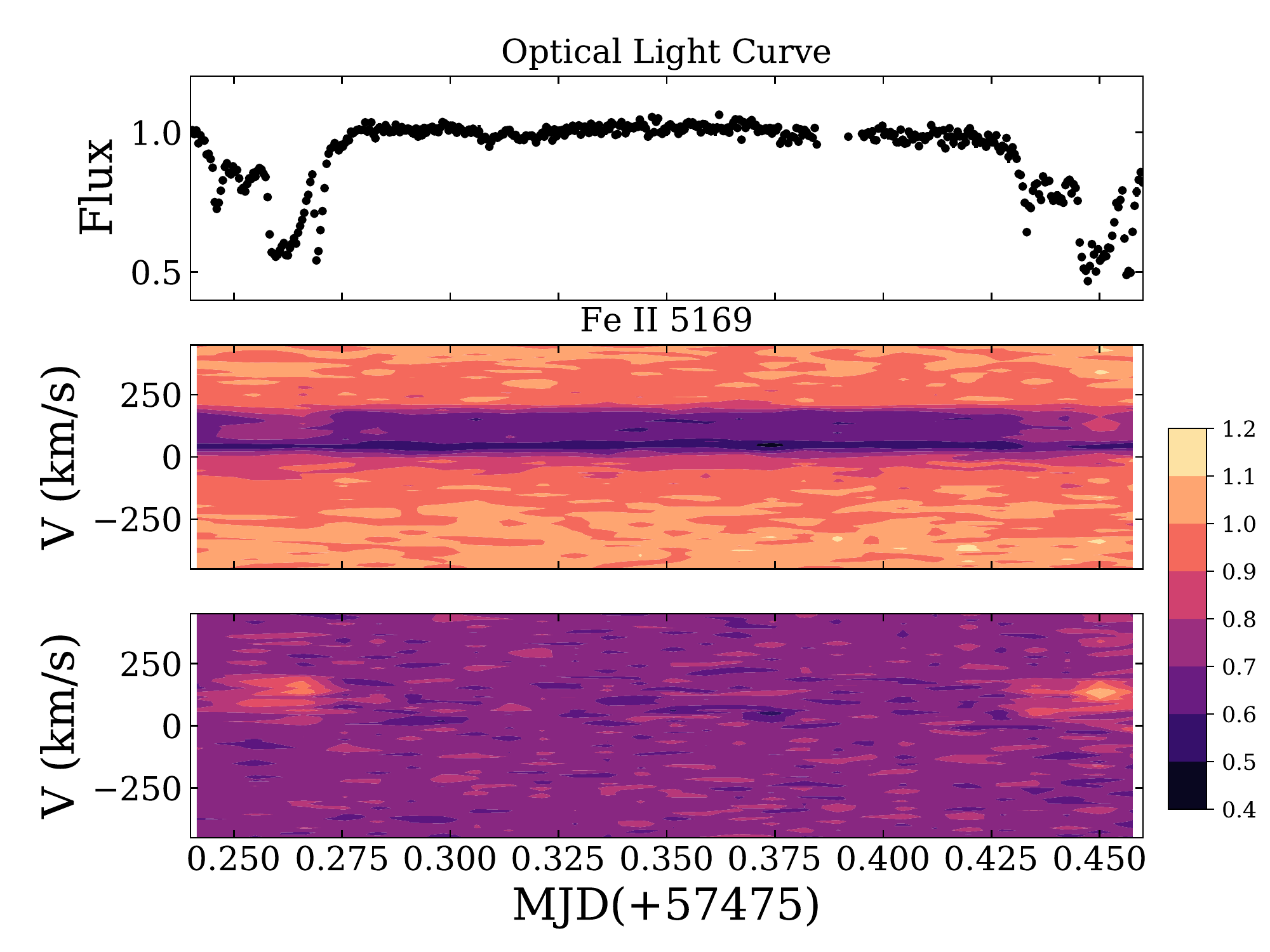}
\caption{Simultaneous photometric and spectroscopic observations of WD~1145+017 on March 28, 2016. The top panel is the optical light curve taken with the 61-inch telescope at Arizona; the middle panel is the {\it Keck }/ESI spectra centered around Fe II 5169 region; the bottom panel is individual spectra divided by the average spectrum around Fe II 5169. During a transit, the absorption feature becomes shallower.
}
\label{fig:lc_spec1}
\end{figure}

\begin{figure*}
\includegraphics[width=\textwidth]{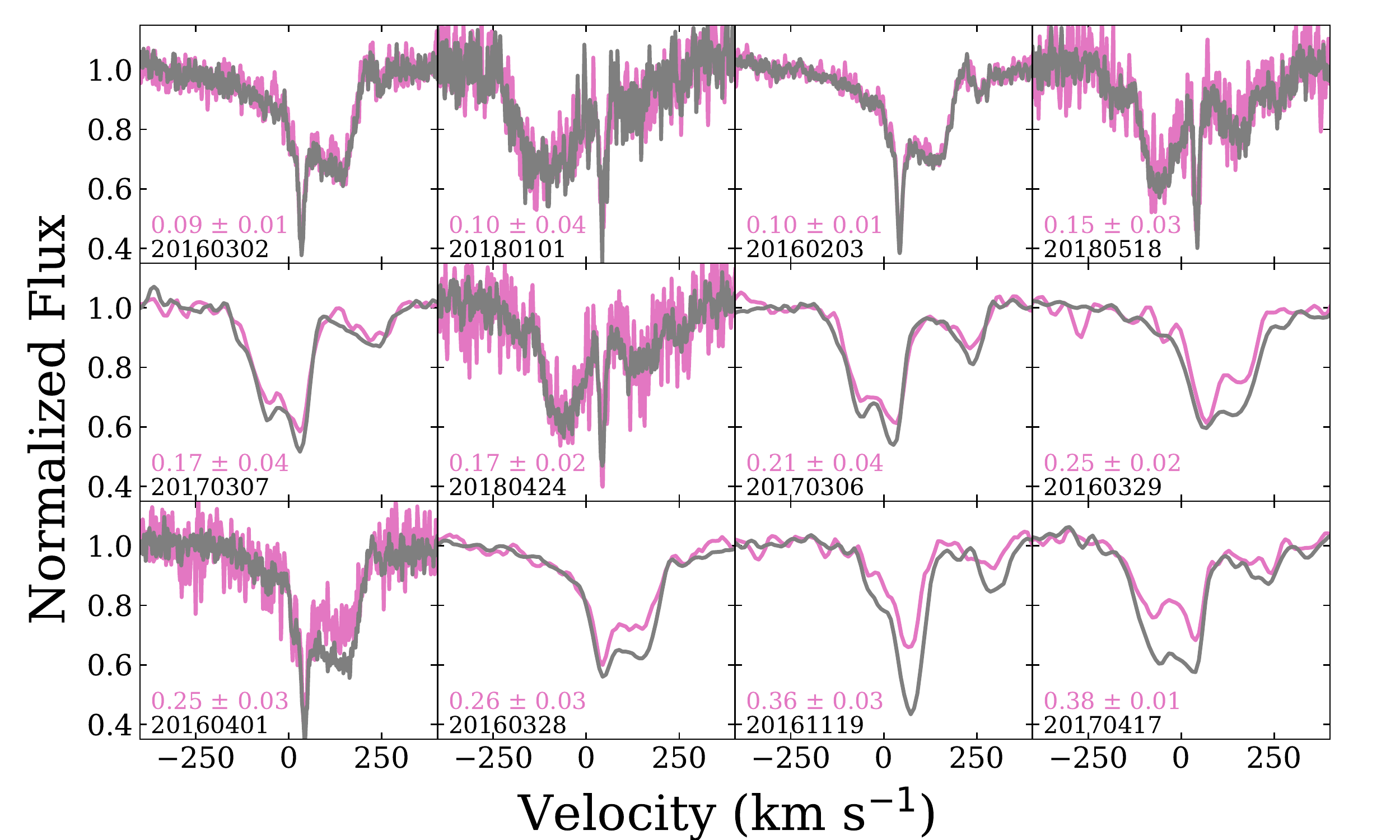}
\caption{In and out-of-transit spectra around Fe II 5169 {\AA}. The in-transit spectra were shown in pink while out-of-transit spectra where shown in grey. The average optical transit depth $\overline{D}$ was listed in pink. From top left to bottom right, the panels were arranged in increasing average transit depth $\overline{D}$. When there is a deep transit, circumstellar lines become shallower, while photospheric lines remain the same.
}
\label{fig:spec_in_out}
\end{figure*}

There are several epochs where we have continuous spectroscopic observations that cover a deep transit, 
suitable for assessing short-term variability of the absorption lines. An example is shown in 
Fig.~\ref{fig:lc_spec1}. During the transit, the absorption feature becomes weaker. Thanks to the continuous optical monitoring of WD~1145+017, we were able to design our spectroscopic observations such that they cover both in-transit and 
out-of-transit spectra, as shown in Fig.~\ref{fig:spec_in_out}. We found that circumstellar lines become 
shallower during a transit but there is little change in photospheric line strength. This effect is most prominent when 
there is a deep transit. Previous spectroscopic studies were around the time when the circumstellar line was mostly 
red-shifted and only a reduction of line strength in the red-shifted component has been reported 
\citep{Redfield2017, Izquierdo2018, Karjalainen2019} and, this study reports the same characteristics apply to the blue-shifted component.

\begin{figure*}
\gridline{\fig{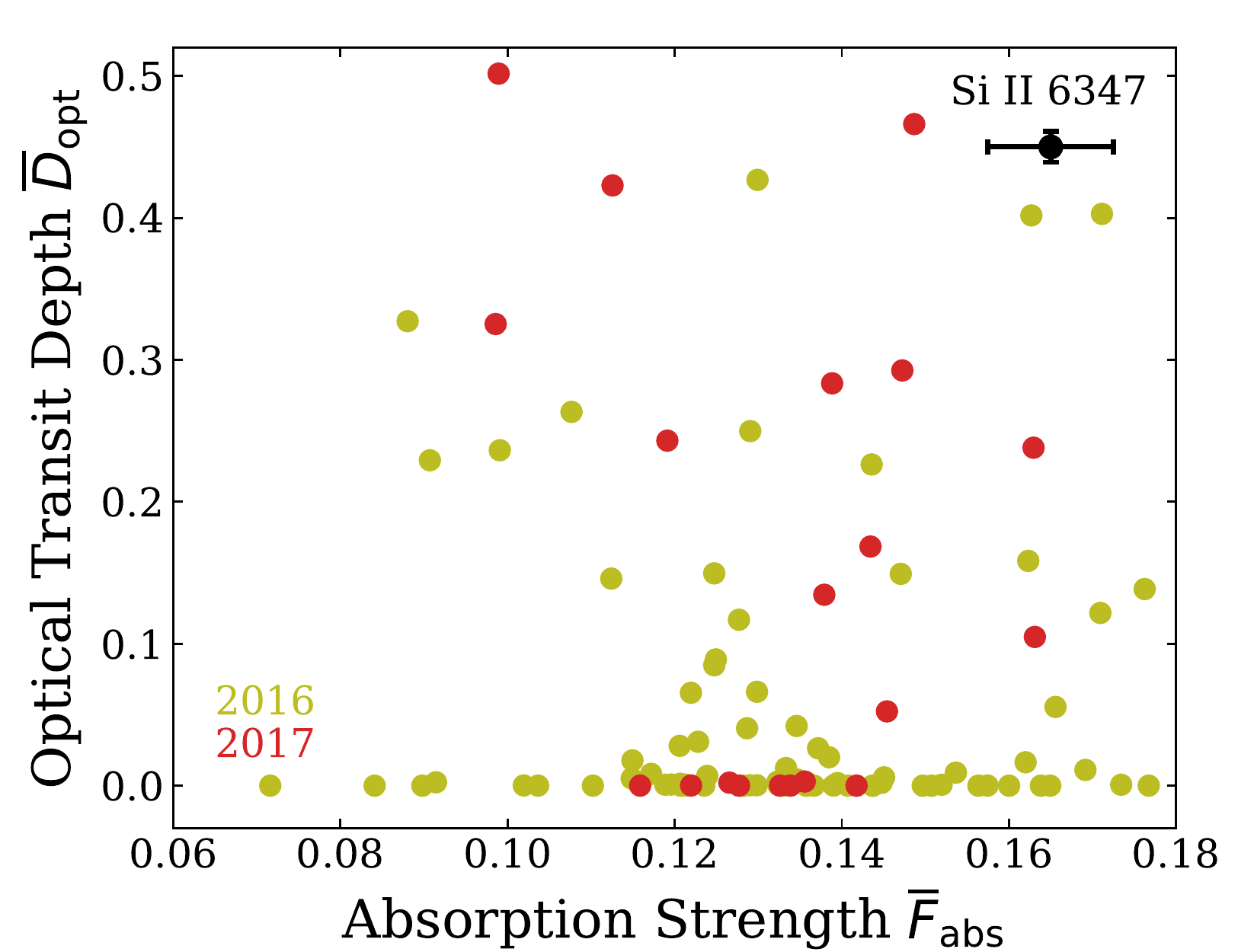}{0.5\textwidth}{}
\fig{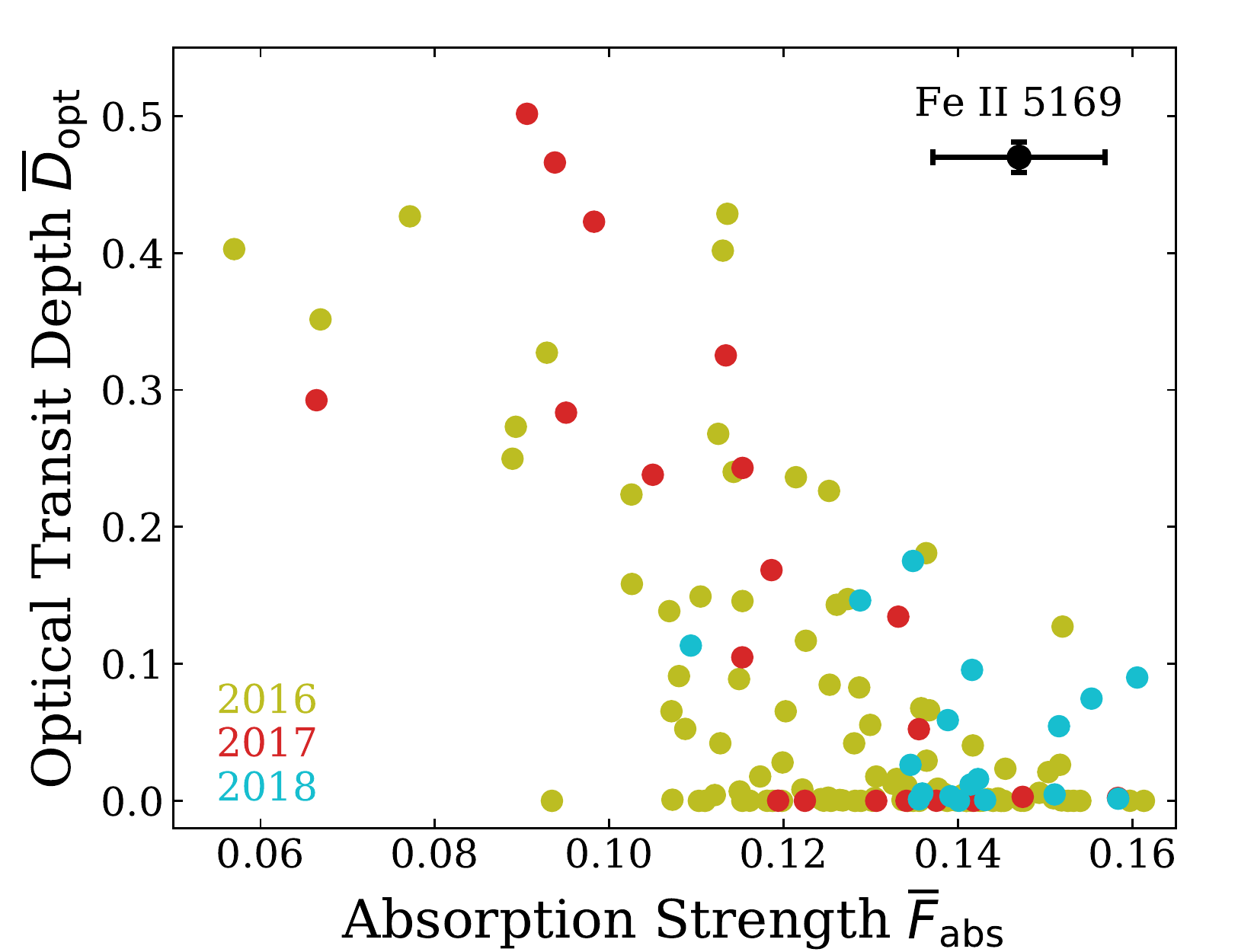}{0.5\textwidth}{} }
\caption{Absorption line strength for individual exposure as a function of transit depth for Si II 6347~{\AA} (102 data points) and Fe 
II 5169~{\AA} (141 data points). The dots are color-coded by the year of the observation. The typical error bar is shown in the upper right corner of each panel. There is a strong anti-correlation between the absorption line strength of Fe II 5169~{\AA} and the 
optical transit depth with a Pearson's correlation coefficient of -0.68, while such a correlation does not exist for the Si II 6347~{\AA} with a correlation coefficient of -0.06.
\label{fig:comp_lc_spec}}
\end{figure*}

As discussed in section~\ref{sec:long-term}, the average strength of the absorption lines (both photospheric 
and 
circumstellar) has remained the same for the past few years. Now we can use every single exposure to assess any possible correlation between the transit depth and the absorption line strength. We focus again on Fe II 5169~{\AA} and Si II 6347~{\AA}. For every spectrum, we calculated the absorption line strength $\overline{F}_\mathrm{abs}$ using Equ.~(\ref{equ:Fabs}) as well as the average optical transit depth $\overline{D}_\mathrm{opt}$ from the optical light curve during the time interval when the spectrum was taken using Equ.~(\ref{equ:D})\footnote{Typically, the spectroscopic and photospheric observations were taken within one night.}. The result is shown in Fig.~\ref{fig:comp_lc_spec}. There is a strong anti-correlation between the line strength of Fe II 5169~{\AA} and the optical transit depth, while the absorption line strengths do not vary much for Si II 6347~{\AA}. We performed the same analysis around other absorption features and found a similar pattern: no short-term variability around absorption lines with only a photospheric component and an anti-correlation between line strength and transit depths among lines with both photospheric and circumstellar contributions.

\section{Interpretation \label{sec:interp} }

\begin{figure*}
\includegraphics[width=\textwidth]{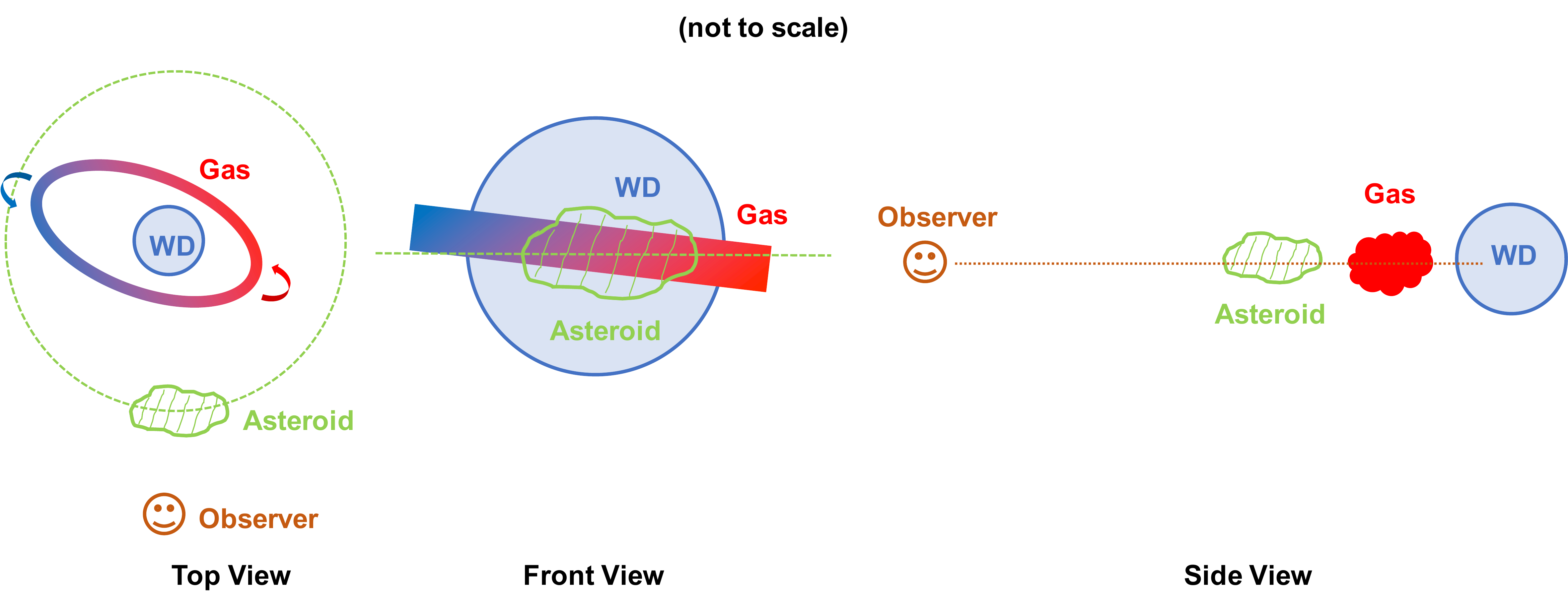}
\caption{A cartoon illustration of our proposed configuration. The orbital plane of the gas disk is aligned 
with that of the dusty fragment. During a transit, the fragment is blocking both the white dwarf and the 
circumstellar gas but with different fractions. \label{fig:cartoon}}
\end{figure*}

Now we explore a toy model that can explain the interplay between circumstellar line strength and transit 
depth. With an orbital period of 4.5~hr, the dusty fragment has a semi-major axis of 
90R$_\mathrm{WD}$. The circumstellar lines can be modeled by a series of eccentric gas rings and the large line width suggests that most of the gas is located close to the white dwarf (e.g. 20-30R$_\mathrm{WD}$ in \citealt{Cauley2018}), within the orbit of the transiting fragment. In addition, our 
model assumes the fragment and the circumstellar gas to be co-planar. A cartoon illustration of our proposed 
geometry is shown in Fig.~\ref{fig:cartoon}.

The transiting fragments are $\sim$ 1000~K and they emit negligible amount of flux at the UV and optical. The out of transit flux $F_\mathrm{out}$ can be calculated as:

\begin{equation}
F_\mathrm{out} = F_\mathrm{WD} - \overline{F}_\mathrm{phot} - \overline{F}_\mathrm{cs}, 
\end{equation}
where the definitions of $F_\mathrm{WD}$, $\overline{F}_\mathrm{phot}$, and $\overline{F}_\mathrm{cs}$ are the same as in Section~\ref{sect:spectra}. Depending on the temperature, circumstellar gas could have some emission as well. $\overline{F}_\mathrm{cs}$ represents the combined flux for the circumstellar gas. From the observed spectra, we know the net result is absorption. 

During a transit, the dusty fragment can block different fractions of the white dwarf and the 
circumstellar gas. We introduce two new parameters $\alpha$ and $\beta$: $\alpha$ characterizes the fraction of white dwarf flux visible
during a transit while $\beta$ characterizes the fraction of circumstellar absorption visible during a transit.
The flux during a transit can be calculated as:

\begin{equation}
F_\mathrm{in} = \alpha \times (F_\mathrm{WD} - \overline{F}_\mathrm{phot}) - \beta \times \overline{F}_\mathrm{cs} 
\end{equation}
If the circumstellar material covers uniformly the whole white dwarf, $\alpha$ would be equal to $\beta$ during a transit.

We can calculate the average transit depth as:
\begin{eqnarray} \label{equ:D2}
\overline{D} &=& \frac{\int (F_\mathrm{out} - F_\mathrm{in}) dp}{\int F_\mathrm{out} dp} \nonumber \\
&=& 1 - \frac{\int^{p_\mathrm{2}}_{p_\mathrm{1}} [ \alpha (p) \times (F_\mathrm{WD} - \overline{F}_\mathrm{phot}) - \beta (p) \times \overline{F}_\mathrm{cs} ] dp}{[F_\mathrm{WD} - \overline{F}_\mathrm{phot} - \overline{F}_\mathrm{cs}] \times (p_\mathrm{2} - p_\mathrm{1})}   \nonumber \\
&=& 1- \frac{ \bar{\alpha} \times (F_\mathrm{WD} - \overline{F}_\mathrm{phot}) - \bar{\beta} \times \overline{F}_\mathrm{cs} }{F_\mathrm{WD} - \overline{F}_\mathrm{phot} - \overline{F}_\mathrm{cs}} 
\end{eqnarray}
where $\bar{\alpha}$ $\equiv$ $\int^{p_\mathrm{2}}_{p_\mathrm{1}} \alpha (p) dp$/ ($p_\mathrm{2} - p_\mathrm{1})$ and $\bar{\beta}$ $\equiv$ $\int^{p_\mathrm{2}}_{p_\mathrm{1}} \beta (p) dp$/ ($p_\mathrm{2} - p_\mathrm{1})$. For a given phase interval ($p_\mathrm{2} - p_\mathrm{1}$), $\bar{\alpha}$ represent the average fraction of detectable light from the white dwarf and $\bar{\beta}$ represents the fraction of visible circumstellar gas. They are geometrical parameters and independent of wavelength. The average transit depth $\overline{D}$ depends on $\bar{\alpha}$, $\bar{\beta}$, $F_\mathrm{WD}$, $F_\mathrm{phot}$, and $F_\mathrm{cs}$. 

In the optical, F$_\mathrm{phot}$ $\ll$ F$_\mathrm{wd}$ and F$_\mathrm{cs}$ $\ll$ F$_\mathrm{wd}$ (see Table~\ref{tab:strength_abs}). We can simplify Equ.~(\ref{equ:D2}) as:

\begin{equation}
\overline{D}_\mathrm{opt} \approx 1- \bar{\alpha}.
\label{equ:ratio_opt}
\end{equation}
The average optical transit depth $\overline{D}_\mathrm{opt}$ has been measured in Table~\ref{tab:transit_par}, and we can calculate the corresponding $\bar{\alpha}$.

In the UV, F$_\mathrm{phot}$ and F$_\mathrm{cs}$ are comparable to F$_\mathrm{wd}$ so we need to use Equ.~(\ref{equ:D2}):

\begin{equation}
\overline{D}_\mathrm{uv} =1- \frac{ \bar{\alpha} \times (F_\mathrm{WD,uv} - \overline{F}_\mathrm{phot,uv}) - \bar{\beta} \times \overline{F}_\mathrm{cs,uv} }{F_\mathrm{WD,uv} - \overline{F}_\mathrm{phot,uv} - \overline{F}_\mathrm{cs,uv}} 
\label{equ:ratio_uv}
\end{equation}

Table~\ref{tab:transit_par} can be used to calculate $\overline{D}_\mathrm{uv}$ and $\bar{\alpha}$. 
$F_\mathrm{WD,uv}$, $F_\mathrm{phot,uv}$, and $F_\mathrm{cs,uv}$ have been reported in 
Table~\ref{tab:strength_abs} in the column labeled COS. Now we can solve for $\bar{\beta}$ using 
Equ.~(\ref{equ:ratio_uv}) and the 
results are also listed in Table~\ref{tab:transit_par}. We see that $\bar{\alpha}$ and $\bar{\beta}$ are 
often unequal; this 
can be explained as the dusty fragment blocking different fractions of the circumstellar gas and the 
white dwarf light. In addition, both $\bar{\alpha}$ and $\bar{\beta}$ varied for different dips, suggesting that the coverages of dusty fragment over circumstellar gas and white dwarf are different.

In Fig.~\ref{fig:comp_lc_spec}, we explored an anti-correlation between the optical transit depth, 
$\overline{D}_\mathrm{opt}$, and absorption strength, $\overline{F}_\mathrm{abs}$. Spectroscopically, 
$\bar{\beta}$ can be calculated as:

\begin{equation} \label{equ:beta2}
\bar{\beta}= \frac{\overline{F}_\mathrm{abs,in} - \overline{F}_\mathrm{phot,in}}{\overline{F}_\mathrm{abs,out} - \overline{F}_\mathrm{phot,out}}.
\end{equation}

\begin{figure} 
\includegraphics[width=0.45\textwidth]{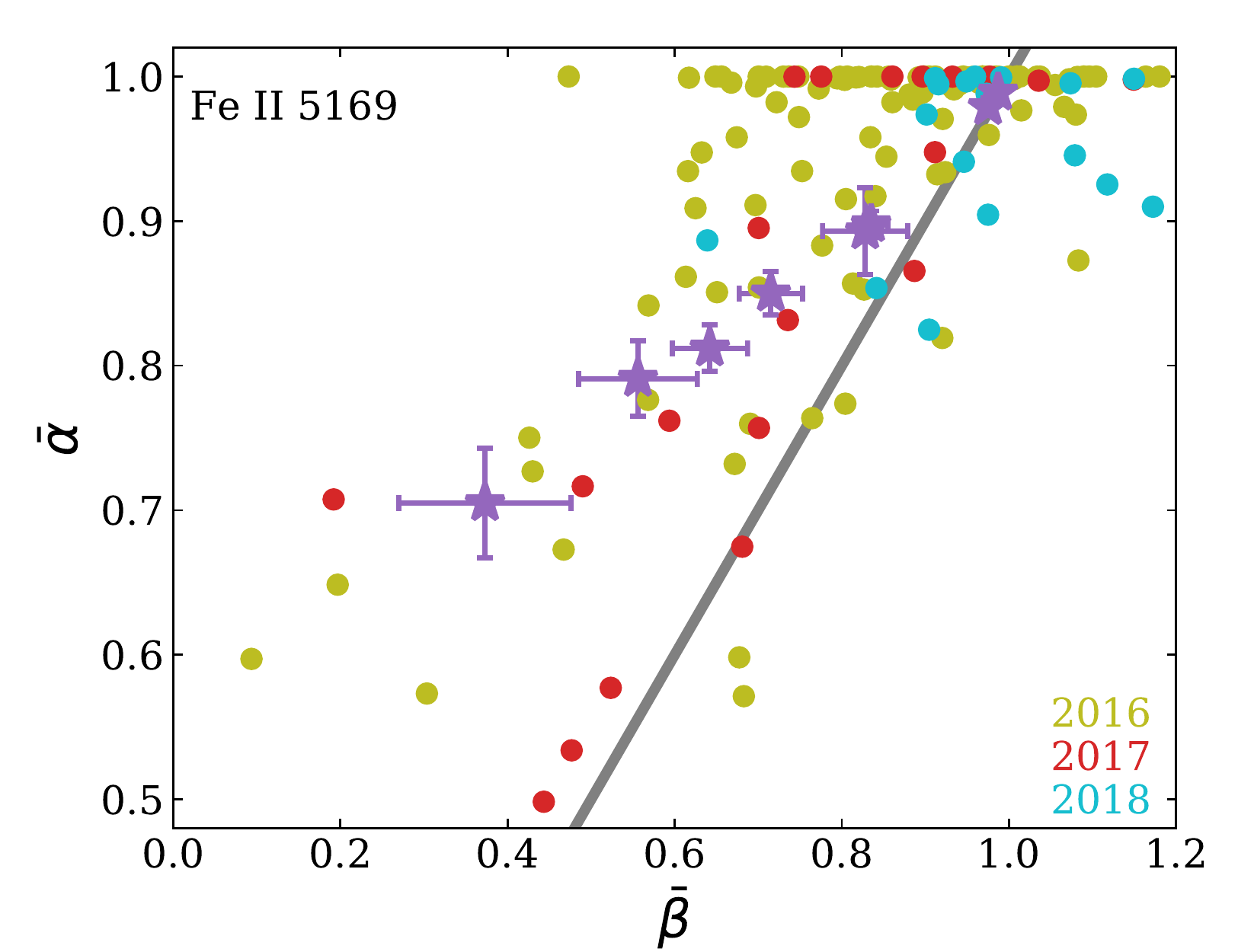}
\caption{This figure is to show the correlation between $\bar{\alpha}$ (the average fraction of white dwarf flux visible during a transit) and $\bar{\beta}$ (the average fraction of circumstellar absorption visible during a transit).  $\bar{\alpha}$ can be directly measured from the optical light curve. For  $\bar{\beta}$, the colored dots are from spectroscopic measurements shown in Fig.~\ref{fig:comp_lc_spec} using Equ.~\ref{equ:beta2}, while the purple stars are from photometric measurement with Equ.~\ref{equ:ratio_uv}. Both sets of measurements follow the same trend and $\bar{\beta}$ is always smaller than $\bar{\alpha}$. The 1:1 ratio line is shown in grey.  \label{fig:alpha_beta2}
}
\end{figure}

We have shown in section~\ref{sec:long-term} and \ref{sec:short-term} that the photospheric line strength is 
constant ($\overline{F}_\mathrm{phot,in}$=$\overline{F}_\mathrm{phot,out}$). Around Fe II 5169, 
$\overline{F}_\mathrm{abs,out}$ = 0.144$\pm$0.018 (from Fig.~\ref{fig:spectra_time}) and 
($\overline{F}_\mathrm{cs}$/$\overline{F}_\mathrm{phot}$) $\approx$ 2 (from fitting the line profile, {\it Fortin-Archambault et al. in prep}). $\overline{F}_\mathrm{abs,in}$ has been measured for individual epoch in 
Fig.~\ref{fig:comp_lc_spec}. As a result, $\bar{\beta}$ can also be directly calculated for an individual 
spectrum. $\bar{\alpha}$ is related to the optical transit depth through Equ~(\ref{equ:ratio_opt}). We can now 
re-arrange the measurements in Fig.~\ref{fig:comp_lc_spec} to represent $\bar{\alpha}$ and $\bar{\beta}$ in 
Fig.~\ref{fig:alpha_beta2}.

There are two ways to determine $\bar{\beta}$, from spectroscopic measurements (Equ.~\ref{equ:beta2}) and photometric measurements (Equ.~\ref{equ:ratio_uv}). We have shown that both sets of measurements yield a similar result in Fig.~\ref{fig:alpha_beta2}: $\bar{\alpha}$ and $\bar{\beta}$ are correlated. Such a correlation has been hinted in \citealt{Hallakoun2017}. They showed in their Fig.~7 that ``bluing" is most prominent in the deepest transit and is less visible for shallower transits.

As shown in Fig.~\ref{fig:alpha_beta2}, $\bar{\beta}$ is always smaller than $\bar{\alpha}$, indicating that the transiting object blocks a larger 
fraction of the circumstellar gas than the white dwarf flux. This is expected as long as the circumstellar gas 
does not uniformally cover the whole white dwarf surface. Because the circumstellar lines are 
highly concentrated in the UV, we detect much less UV circumstellar absorption during a transit. As a result, 
the observed flux appears to be higher in the UV compared to the optical and therefore the UV-to-optical transit depth ratios are less than unity. This is somewhat in analog to a spot occultation during a transiting planet. Because a starspot is fainter than the surrounding area, the light curve gets a bit brighter when a planet transits across the starspot compared to other parts of the stellar surface. 

This model can quantitatively explain the shallower UV transits observed around WD~1145+017. After correcting for the change of circumstellar lines, the UV-to-optical transit depth ratios are consistent with unity from 2016 to 2018. 

Note that an important conclusion of the model is that the disintegrating object, circumstellar gas, and the 
white dwarf are aligned along our line of sight -- the system is edge-on. This is expected because likely, the circumstellar gas is 
produced by the sublimation or collision of the fragments from the disintegrating objects \citep{Xu2018a}. There could be some gas around the same region of the dusty fragments and this analysis still holds if the circumstellar gas extends beyond the orbit of the dusty fragment. Shallower UV transits are expected as long as the transiting fragment is blocking a larger fraction of the circumstellar gas compared to the white dwarf surface.

\section{Conclusion \label{sec:final}}

In this paper, we report multi-epoch photometric and spectroscopic observations of WD~1145+017, a white dwarf 
with an actively disintegrating asteroid. The main conclusions are summarized as 
follows:

$\bullet$ {\it There is a strong anti-correlation between circumstellar line strength and transit depth.} Regardless 
of being blue-shifted or red-shifted, the circumstellar lines become 
significantly weaker during a deep 
transit. This can be explained when the transiting fragment is blocking a larger fraction of the circumstellar 
gas than the white dwarf flux during a transit.

$\bullet$ {\it The shallow UV transit is a result of short-term variability of the circumstellar lines.} The UV transit depth is always shallower than that in the optical. We presented a model that can quantitatively explain this phenomena due to the deduction of circumstellar line strength during a transit and their high concentration in the UV.

$\bullet$ {\it The orbital planes of the gas disk and the dusty fragment are likely to be aligned}. An important conclusion of our model is the alignment between the transiting fragment and circumstellar gas -- the system is edge-on. This is consistent with the picture that the gas is likely to come from the fragment and is eventually accreted onto the white dwarf.

$\bullet$ {\it We have yet to detect any differences in the transit depths directly caused by the transiting 
material itself.} We have not detected any wavelength dependence caused by the transiting material at wavelengths 
from 0.1~$\mu$m to 4.5~$\mu$m. The transiting material must be either optically thick at all of these 
wavelengths or mostly consist of 
large particles.

One main puzzle left for WD~1145+017 is the location and evolution of the dust disk, which could be constrained by future observations. Infrared spectroscopy could put limits on the temperature, size, and composition of the dust disk while infrared monitoring  can probe the evolution of the disk.

\smallskip
{\it Acknowledgements.} The authors would like to thank S. Rappaport for useful discussions on different aspects of the manuscript. We also greatly appreciate helps from R. Alonso, D. Bayliss, P. Benni, K. Collins, D. Conti, N. Espinoza, P. Evans, E. M. Green, J. Hambsch, T. Kaye, J. Kielkopf, M. Motta, Y. Ogmen, E. Palle, H. Relles, S. Sagear, S. Sharma, A. Shporer, C. Stockdale, T. G. Tan, and G. Zhou for performing optical observations of WD~1145+017, L. Hillenbrand and C. Melis for some {\it Keck}/HIRES observations, V. D. Ivanov and S. Randall for designing {\it VLT} observations, B. Gaensicke for discussing {\it HST} observing strategy, and J. Ely for discussions on extracting COS light curves. N. H. acknowledges the hospitality of the European Southern Observatory in Garching.

The paper was based on observations made with the NASA/ESA Hubble Space Telescope under program 14467, 14646, and 15155, obtained from the data archive at the Space Telescope Science Institute. STScI is operated by the Association of Universities for Research in Astronomy, Inc. under NASA contract NAS 5-26555. This paper also reported observations from the Spitzer Space Telescope under programme no. 13065, which is operated by the Jet Propulsion Laboratory, California Institute of Technology under a contract with National Aeronautics and Space Administration (NASA). There are some data obtained from the European Organisation for Astronomical Research in the Southern Hemisphere under European Southern Observatory (ESO) programme 296.C-5024.

A large amount of data presented herein were obtained at the W. M. Keck Observatory, which is operated as a scientific partnership among the California Institute of Technology, the University of California and the National Aeronautics and Space Administration. The Observatory was made possible by the generous financial support of the W. M. Keck Foundation. The authors wish to recognize and acknowledge the very significant cultural role and reverence that the summit of Maunakea has always had within the indigenous Hawaiian community.  We are most fortunate to have the opportunity to conduct observations from this mountain.

These results also made use of the Discovery Channel Telescope at Lowell Observatory. Lowell is a private, 
non-profit institution dedicated to astrophysical research and public appreciation of astronomy and operates 
the DCT in partnership with Boston University, the University of Maryland, the University of Toledo, Northern 
Arizona University and Yale University.  The Large Monolithic Imager was built by Lowell Observatory using 
funds provided by the National Science Foundation (AST-1005313).  

A portion of this research was supported by 
NASA and NSF grants to UCLA. This work is partly supported by JSPS KAKENHI Grant Numbers JP18H01265, 18H05439 and JP16K13791, and JST PRESTO Grant Number JPMJPR1775.

\end{CJK}

\software{Matplotlib \citep{Matplotlib}}

\end{document}